\documentclass{article}

 \usepackage{graphicx}
\usepackage{bm}
\usepackage{amsfonts}
\usepackage{amsmath}

\usepackage{graphicx}
\usepackage{epsfig}
\usepackage{amsfonts}
\usepackage{a4}
\usepackage{amsmath}

\newcommand{\vphi}{\varphi}

\def\dalemb#1#2{{\vbox{\hrule height .#2pt
        \hbox{\vrule width.#2pt height#1pt \kern#1pt
                \vrule width.#2pt}
        \hrule height.#2pt}}}

\def\0{{\sst{(0)}}}
\def\1{{\sst{(1)}}}
\def\2{{\sst{(2)}}}
\def\3{{\sst{(3)}}}
\def\4{{\sst{(4)}}}
\def\5{{\sst{(5)}}}
\def\6{{\sst{(6)}}}
\def\7{{\sst{(7)}}}
\def\8{{\sst{(8)}}}

\def\wtd{\widetilde}

\let\w=\omega

\def\nn{\nonumber} \def\bd{\begin{document}} \def\ed{\end{document}}
\def\ds{\documentstyle} \let\fr=\frac \let\bl=\bigl \let\br=\bigr
\let\Br=\Bigr \let\Bl=\Bigl
\let\bm=\bibitem
\let\na=\nabla
\let\pa=\partial \let\ov=\overline
\def\fft#1#2{{#1 \over #2}}

\begin{document}

\title{
Non-Abelian fields in AdS$_4$ spacetime:
\\
axially symmetric, composite configurations}

\author{{\large Olga Kichakova}$^{1}$, {\large  Jutta Kunz}$^{1}$,
{\large Eugen Radu}$^{2}$
and {\large Yasha Shnir}$^{1,3,4}$
\\
\\
 {\small  $^{1}$Institut f\"ur Physik, Universit\"at Oldenburg, Postfach 2503
D-26111 Oldenburg, Germany}
\\
 {\small  $^{2}$Departamento de Fisica da Universidade de Aveiro and I3N,
 Campus de Santiago, 3810-183 Aveiro, Portugal}
\\
{\small $^{3}$Department of Theoretical Physics and Astrophysics, BSU, Minsk, Belarus}
\\
{\small $^{4}$BLTP, JINR, Dubna, Russia}
 }
\date{\today}

\maketitle

 \begin{abstract}
We construct new finite energy regular solutions in Einstein-Yang-Mills-SU(2) theory.
They are static, axially symmetric and approach at infinity the anti-de Sitter
spacetime background.
These configurations are characterized by a pair of integers $(m, n)$,
where $m$ is related to the
polar angle and $n$ to the azimuthal angle,
being related to the known flat space
monopole-antimonopole chains and vortex rings.
Generically, they describe
composite configurations with several individual components,
possesing a nonzero magnetic charge,
even in the absence of a Higgs field.
Such Yang-Mills  configurations exist already in the probe limit,
the AdS geometry supplying the attractive force needed to balance
the repulsive force of Yang-Mills gauge interactions. 
The gravitating  solutions are constructed
 by numerically solving the elliptic Einstein-DeTurck--Yang-Mills equations. The
variation of the gravitational coupling constant $\alpha$ reveals the existence of two
branches of gravitating solutions which bifurcate at some critical value of $\alpha$.
The lower  energy branch connects to the solutions in the global AdS spacetime,
while the upper branch
is linked to the generalized Bartnik-McKinnon solutions in asymptotically flat spacetime.
%
Also, a spherically symmetric, closed form
solution is found as a perturbation around the globally anti-de Sitter vacuum state.
 \end{abstract}

\section{Introduction}

The study of solutions of the Yang-Mills (YM) theory
in a curved spacetime geometry can be traced back at least to the early work
\cite{BoutalebJoutei:1979va}.
Among other results, that study has given an exact solution of the
YM equations in a fixed Schwarzschild black hole background.
This shows that the non-trivial solutions to the
full system of Einstein--Yang-Mills (EYM) equations are likely
to exist, at least for large enough event horizon black hole radius.

Indeed, this has been confirmed ten years later, when several different authors
have constructed asymptotically flat,
black hole (BH) solutions within the framework of $d = 4$  SU(2)   EYM  theory \cite{Volkov:1989fi}.
Although these BHs were static and spherically symmetric, with
vanishing  YM  charges, they were different from the Schwarzschild one and,
therefore, not characterized exclusively by their total mass.
Unfortunately, soon after their discovery, it has
been shown that these solutions are unstable
\cite{Volkov:1994dq},
\cite{Zhou:1991nu}.
However, despite this fact, they still present a challenge to the
standard {\it 'no hair conjecture'}
\cite{Wheeler},
\cite{Bekenstein:1996pn}.

These results have led to a revision  of some of the basic
concepts of BH physics based on the uniqueness and no-hair theorems.
For example, the Israel�s theorem does not generalize to
the non-Abelian (nA) case, since static EYM black
holes with non-degenerate horizon turn
out to be not necessarily spherically
symmetric \cite{Kleihaus:1997ic}.
Moreover,
in strong contrast to the Abelian case, the EYM hairy black holes in  \cite{Volkov:1989fi}
do not trivialize in the limit of a vanishing horizon
area\footnote{In fact, rather curious,
the EYM particle-like solutions
have been discovered \cite{Bartnik:1988am} before their black hole generalizations.},
reducing to horizonless, globally regular, particle-like configurations,
originally
found by Bartnik and McKinnon in Ref. \cite{Bartnik:1988am}.

As a result,
the subject of particle-like and hairy BH solutions in EYM theory has become an active field of research,
with many new results being reported each year\footnote{
A detailed review of the situation ten years after the discovery of the EYM solutions in
\cite{Volkov:1989fi},
\cite{Bartnik:1988am}
can be found in \cite{Volkov:1998cc}.}.
One interesting question addressed in this context is what happens if one drops
the assumption of asymptotic flatness for the spacetime background.
The case of the EYM system with a negative cosmological constant $\Lambda$ is of particular interest,
the natural background of the theory
corresponding to anti-de Sitter (AdS) spacetime.
Solutions of various physical models in this geometry
received recently much interest due to the conjectured
 anti-de Sitter/Conformal Field Theory (AdS/CFT) correspondence.
 This is
 a concrete realization of the holographic principle, which asserts that a consistent
theory of quantum gravity in $d$-dimensions must have an alternate formulation in terms
of a nongravitational theory in $(d-1)$-dimensions.
Despite the fact that string theory in AdS space is still too
complicated to be dealt with in detail,
in many interesting cases, it is sufficient to consider
the low energy limit of the superstring theory, namely, supergravity.
However, the gauged supergravity theories generically contain
 YM fields (although most of the studies in the literature have been restricted to the case of Abelian
matter content in the bulk),
and thus the interest in the study of the EYM system with $\Lambda<0$.

The first result on nA fields in a globally
AdS$_4$ geometry can be found again in Ref.
\cite{BoutalebJoutei:1979va}, where a non-trivial solution
of the YM equations is exhibited in closed form\footnote{However, note that the Ref. \cite{BoutalebJoutei:1979va} has considered only the
case of a positive cosmological constant and a slightly different coordinate system as compared to (\ref{AdS}).}.
This solution describes a globally regular, finite energy soliton,
with a nonzero magnetic flux at infinity (despite the absence of a Higgs field),
anticipating most of the basic properties of the
(gravitating) nA fields in an AdS background.

The study of  EYM
solutions in a globally AdS$_4$ background have started with Refs.
\cite{Winstanley:1998sn}, \cite{Bjoraker:2000qd}, where spherically symmetric
BHs and solitons have been studied, again for the gauge group SU(2).
As shown there,
a variety of well known features of asymptotically flat self-gravitating nonabelian
solutions are not shared by their AdS counterparts.
Restricting for simplicity to purely magnetic configurations,
one finds a continuum of particle-like and BH
solutions describing nA monopoles
with a non-integer magnetic charge
(we recall that the asymptotically flat EYM configurations are magnetically neutral,
forming a discrete sequence indexed by the node number of the magnetic gauge potential \cite{Volkov:1989fi}).
Moreover, perhaps most remarkable, some of the solutions
are stable against spherically symmetric
linear perturbations
\cite{Sarbach:2001mc},
\cite{Breitenlohner:2003qj}.
Also, as already found in Ref. \cite{BoutalebJoutei:1979va},
the curved background geometry provided an extra attracting force,
which makes possible the existence of finite mass, particle-like YM solutions
already in the probe limit,
$i.e.$ in a fixed AdS spacetime.
As discussed in
\cite{Baxter:2007at},
\cite{Baxter:2007au}
the
soliton and black hole solutions in \cite{Winstanley:1998sn}, \cite{Bjoraker:2000qd}
possess interesting generalizations with higher gauge groups.
A review of the EYM solutions in a globally AdS$_4$ background can be found in Ref. \cite{Winstanley:2008ac}.

We note that an even more intricate picture is found when
studying EYM  topological  BHs \cite{VanderBij:2001ia}.
For example, no globally regular particle-like limit of the solutions
is found in this case.
Moreover,
as discovered in \cite{Gubser:2008zu},
 \cite{Gubser:2008wv}, the planar hairy BHs
 describe gravity duals of $p-$wave superconductors.
 As a result,
 this type of EYM solutions enjoyed recently much interest.
Also, the $d=4$ configurations possess higher dimensional generalizations  \cite{Manvelyan:2008sv},
with
the $d=5$  EYM  planar black holes describing holographic $p-$wave superfluids \cite{Ammon:2009xh}.
However, these aspects are beyond the purposes of this paper,
where we shall restrict ourselves to the case of the
SU(2) gauge fields and the asymptotically  AdS$_4$ spacetime.
%
One should remark that, even in the case of the global AdS spacetime,
the study of YM solutions is still far
from being complete.
For example, very few things are known about non-spherically symmetric
EYM-AdS solutions.
Non-Abelian solitons which are axially symmetric
only have been studied\footnote{Their black hole generalizations have been considered in
  \cite{Radu:2004gu}.} in Ref.  \cite{Radu:2001ij}.
 These configurations possess an azimuthal winding number $n>1$
 and describe (non-topological) 
 monopoles localized at the origin,
 sharing all basic properties
of the spherically symmetric counterparts (which have $n=1$).
Also, despite the generic
presence of a net magnetic flux,
they can be viewed as the natural AdS generalizations of the
asymptotically flat EYM solitons in
\cite{Kleihaus:1997mn}.

However, as discussed in
\cite{Ibadov:2004rt},
the EYM system with $\Lambda=0$
possesses a variety of other globally regular  solutions, describing {\it composite configurations},
with several constituents\footnote{Note that no such solutions exist with Abelian matter fields only,
the closest approximation there being the Majumdar-Papapetrou
\cite{Majumdar:1947eu},
\cite{Papaetrou:1947ib}
extremal black holes in Einstein-Maxwell theory.}.
In the notation of
\cite{Ibadov:2004rt},
these solutions are
characterized by a pair of positive integers $(m, n)$, where $m$ is related to the
polar angle and $n$ to the azimuthal angle.

One should remark that the AdS solutions discussed so far in the literature cover the case $m=1$ only,
that is, solutions with a single center.
The main purpose of this work is to explicitly
construct AdS $composite$ configurations with $m>1$,
looking for new features induced
by the different asymptotic structure of the spacetime.

Although some common features are present, the results we find for $\Lambda<0$ are rather different from those valid
in the asymptotically flat case.
Perhaps the most prominent new result is the existence of
 multi-center solutions with a net magnetic
flux at infinity, despite the absence of a Higgs field
(such solutions are absent
 for $\Lambda=0$).
Similar to the spherically symmetric case \cite{BoutalebJoutei:1979va},
such nA configurations exist
already in the probe limit ($i.e.$ no backrection).
Moreover, when including the gravity effects,
we establish the absence, for large enough values of $|\Lambda|$,
of configurations with a zero magnetic flux
(which are the only ones existing in the asymptotically flat case).

This paper is structured as follows:
 in the next Section we introduce the  gauge field ansatz and
 address the question of possible asymptotics
for (purely magnetic) static,
axially symmetric YM fields.
In Section 3 we construct solutions with these asymptotics
in the probe limit, $i.e.$
for a fixed  AdS background.
Although being relatively simple, nevertheless this case appears to contain all the essential features
of the gravitating solutions.
The backreaction of the solutions on the spacetime geometry is
considered in Section 4.
There special attention is paid to  a particular value of $\Lambda$,
for which the EYM system becomes a consistent truncation of the
$d=4$, ${\cal N}=4$ gauged supergravity \cite{Cvetic:1999au}; therefore such
solutions can be uplifted to $d=11$ dimensions  \cite{Pope:1985bu}.
When we abandon this restriction, the variation of the gravitational coupling constant $\alpha$
(which is the ratio between the Planck length and AdS length scales) 
reveals the existence of two
branches of gravitating solutions which bifurcate at some critical value $\alpha_{cr}$.
These branches interpolate between the solutions in the global AdS spacetime and,
for small values of the coupling constant on the upper branches,
the generalized Bartnik-McKinnon solutions in the asymptotically flat spacetime \cite{Ibadov:2004rt},
confined in the interior region, and outer configurations in the global AdS spacetime.

We give our conclusions and remarks in the final Section.
The Appendix A contains a derivation of an exact
solution of the EYM equations with negative cosmological constant as a perturbation around the globally AdS
vacuum state.

\section{SU(2) Yang-Mills fields on AdS$_4$}

\subsection{The model}
We consider the action of the  SU(2)  YM
theory  \begin{eqnarray}
\label{model}
S= -\frac{1}{2} \int d^4 x\sqrt{-g}
{\rm Tr}
\big \{
F_{\mu\nu}F^{\mu\nu}
 \big \},
\end{eqnarray}
 with the
 field strength tensor
\begin{equation}
F_{\mu\nu} = \partial_\mu A_\nu - \partial_\nu A_\mu + i e [A_\mu, A_\nu] \ ,
\end{equation}
and  the gauge potential
\begin{equation}
A_\mu= \frac{1}{2}\tau_a A_\mu^a,
\end{equation}
 $e$  being the  gauge coupling constant.
Also, $\mu,\nu$ are
 space-time indices running from 1 to 4 and the gauge index $a$ is running from 1 to 3.


Variation of   (\ref{model}) with respect to the gauge field $A_\mu$
leads to the YM equations
\begin{eqnarray}
\label{feqA}
  D_\mu F^{\mu\nu}
  \equiv
  \nabla_\mu F^{\mu\nu}+i e[A_\mu, F^{\mu\nu} ]
  =0,
\end{eqnarray}
while the variation with respect to the metric $g_{\mu\nu}$ yields the energy-momentum tensor
of the YM fields
 %
\begin{eqnarray}
\label{Tik}
&T_{\mu\nu} =
  2 {\rm Tr}
   \big \{
      F_{\mu\alpha} F_{\nu\beta} g^{\alpha\beta}
   -\frac{1}{4} g_{\mu\nu} F_{\alpha\beta} F^{\alpha\beta}
   \big \}
\ .
\end{eqnarray}

 For the background metric, we shall consider the (covering-)AdS$_4$  spacetime, written in global coordinates as
\begin{eqnarray}
\label{AdS}
ds^2=\frac{dr^2}{N(r)}+r^2(d\theta^2+\sin^2\theta d\varphi^2)-
N(r)dt^2,~~{\rm with}~~N(r)=1+\frac{r^2}{\ell^2},
 \end{eqnarray}
where $(r,t)$ are the radial and time coordinates, respectively  (with $0\leq r<\infty$ and $-\infty<t<\infty$), while
$\theta$ and $\varphi$ are angular coordinates  with the usual range, parametrizing the two dimensional sphere $S^2$.
Also, $\ell$ is the AdS length scale,
 which is fixed by the cosmological constant,
\begin{eqnarray}
\label{ell}
\Lambda=-\frac{3}{\ell^2}.
\end{eqnarray}

\subsection{The axially symmetric YM Ansatz}
 In this work
we shall restrict to purely magnetic YM configurations
and  employ a gauge field ansatz in the parametrization\footnote{This is in fact a suitable reparametrization of
the axially symmetric YM ansatz introduced for the first time
 by Manton \cite{Manton:1977ht}
 and Rebbi and Rossi \cite{Rebbi:1980yi},
which is better suited for numerical purposes.
Also, note that (\ref{gauge-ansatz}) is a consistent truncation of the most general
YM ansatz, which contains 12 potentials.
}
originally proposed in \cite{Kleihaus:1997mn}
 \begin{eqnarray}
\label{gauge-ansatz}
 A_\mu dx^\mu=
\left( \frac{H_1}{r} dr + (1-H_2)d\theta\right)\frac{u_\vphi^{(n)}}{2e}
- n \sin\theta \left( H_3\frac{u_r^{(n)}}{2e}
                     + (1-H_4)\frac{u_\theta^{(n)}}{2e}\right) d\vphi,
\end{eqnarray}
in terms of four gauge field functions $H_i$ which depend on $r$ and $\theta$ only.
The SU(2) matrices $u_a^{(n )}$ factorize the dependence on the azimuthal coordinate $\varphi$,
with
\begin{eqnarray}
\nonumber
u_r^{(n )}&=&\sin  \theta (\cos n\varphi~\tau_x+\sin n\varphi~\tau_y)+\cos \theta ~\tau_z,
\\
\nonumber
u_\theta^{(n )}&=&\cos \theta (\cos n\varphi~\tau_x+\sin n\varphi~\tau_y)-\sin \theta ~\tau_z,
\\
\nonumber
u_\varphi^{(n)}&=&-\sin n\varphi~\tau_x+\cos n\varphi~\tau_y ,
\end{eqnarray}
where 
 $ \tau_x, \tau_y, \tau_z $ are the Pauli matrices.
The positive integer $n$ represents the  azimuthal winding number of the solutions.
For $n=1$ and $H_1=H_3 =0$, $H_2=H_4=w(r)$
the usual spherically symmetric magnetic (singularity-free) YM ansatz
is recovered.

This ansatz is axially symmetric in the sense that a rotation
around the $z-$axis (with $z=r\cos\theta$) can be compensated  by a suitable gauge transformation
\cite{Heusler:1996ft,Forgacs:1980zs}.
However, note that
the gauge transformation
$
U = \exp \{i \Gamma (r,\theta) u_\vphi^{(n)}/2\}
$
leaves the ansatz form-invariant \cite{Brihaye:1994ib}.
Thus, to construct regular solutions we have to fix the gauge.
The usual
gauge condition
\cite{Kleihaus:1997mn}
 which is used also in this work is
\begin{eqnarray}
\label{gauge}
r \partial_r H_1 - \partial_\theta H_2 = 0.
\end{eqnarray}

A straightforward computation leads to the following expression of
the non-vanishing components of the SU(2) field strength tensor
(with $F_{\mu\nu}=F_{\mu\nu}^{(a)}\frac{1}{2e}u_{a}^{(n)}$):
\begin{eqnarray}
\label{Fij}
&&
F_{r\theta}^{(\varphi)}=-\frac{1}{r}(H_{1,\theta}+rH_{2,r}),
\\
&&
\nonumber
F_{r\varphi}^{(r)}=-\frac{n\sin\theta}{r}(r H_{3,r}-H_1H_{4}),
~~
F_{r\varphi}^{(\theta)}= \frac{n\sin\theta}{r}(r H_{4,r}+H_1 H_{3}+\cot\theta H_1),
\\
&&
\nonumber
F_{\theta\varphi}^{(r)}= -n\sin\theta ( H_{3,\theta}-1+H_2 H_{4}+\cot\theta H_3),
~~
F_{\theta\varphi}^{(\theta)}= n\sin\theta ( H_{4,\theta}-H_2 H_{3}-\cot\theta (H_2-H_4) ),
\end{eqnarray}

We are interested in singularity-free solutions of the YM equations (\ref{feqA})
with a finite mass.
For configurations in a fixed AdS background,
the total mass $M$ is defined as the integral of the
mass-energy density $\rho=-T_t^t$ over a $t=const.$ three-dimensional space, $i.e.$
%
\begin{eqnarray}
\label{mass}
M=-\int d^3 x\sqrt{-g}T_t^t=-2\pi \int_0^\infty dr \int_0^{\pi} d\theta~r^2 \sin \theta_{~} T_t^t.
 \end{eqnarray}
From (\ref{Tik}) and (\ref{gauge-ansatz}), one finds
\begin{eqnarray}
\label{Ttt}
-T_t^t=\frac{1}{2e^2r^2}
\bigg (
NF_{r\theta}^2+\frac{1}{\sin^2\theta}(NF_{r\varphi}^2+\frac{1}{r^2}F_{\theta\varphi}^2 )
\bigg),
 \end{eqnarray}
 where we denote
\begin{eqnarray}
F_{r\theta}^2= (F_{r\theta}^{(\varphi)})^2 ,~~
F_{r\varphi}^2= (F_{r\varphi}^{(r)})^2 +(F_{r\varphi}^{(\theta)})^2,~~
 F_{\theta\varphi}^2= (F_{\theta\varphi}^{(r)})^2 +(F_{\theta\varphi}^{(\theta)})^2.
\end{eqnarray}

As known already from the study of   spherically symmetric
configurations \cite{Bjoraker:2000qd},
a generic feature of the generic YM solutions in an AdS$_4$ background is that they
 possess a nonvanishing magnetic flux
through the sphere at infinity.
A measure of this flux is provided by the magnetic charge, $Q_M$, which, in the absence of the Higgs field,
does not have a meaning of a topological charge of the configuration, thus it is allowed to be non-integer.
A possible gauge invariant definition of $Q_M$
which we shall employ in this work, is
 \cite{Corichi:2000dm}
\begin{eqnarray}
\label{Qm}
Q_M=\frac{1}{4\pi}\oint_\infty d\theta d\varphi \sqrt{ (F_{\theta\varphi}^{(r)})^2 +(F_{\theta\varphi}^{(\theta)})^2}.
\end{eqnarray}
We have verified that in the spherically symmetric case,
(\ref{Qm}) agrees, up to a sign, with the
magnetic charged expression in \cite{Bjoraker:2000qd}.

We close this part by noticing  that the
static axially symmetric YM configurations in a fixed AdS spacetime
satisfy the
virial identity\footnote{As usual, this virial identity is found by considering
the scale transformation $r\to \lambda r$ of the effective action $S_{eff}$
of the model (which essentially coincides with (\ref{mass})),
for a given set of boundary conditions.
Then  $S_{eff}$ must have a critical point at $\lambda=1$,
which results in the virial relation (\ref{virial}).
}
\begin{eqnarray}
\label{virial}
 \int_0^\infty dr \int_0^{\pi} d\theta  \sin \theta{~}
 \bigg (
  NF_{r\theta}^2+\frac{1}{\sin^2\theta}(NF_{r\varphi}^2+\frac{1}{r^2}F_{\theta\varphi}^2 )
  \bigg)
  =
   \int_0^\infty dr \int_0^{\pi} d\theta  \sin \theta_{~}
   \frac{2r^2}{\ell^2}
 \bigg (
 F_{r\theta}^2+\frac{1}{\sin^2\theta} F_{r\varphi}^2
  \bigg).
 \end{eqnarray}
 This makes clear that the AdS geometry
 supplies the attractive force needed to balance
the repulsive force of Yang-Mills gauge interactions.

\subsection{The issue of boundary conditions at infinity}

\subsubsection{$\Lambda=0$ flat space case}
Let us start with the more familiar case of  gauge fields in a Minkowski spacetime background.
For $\Lambda=0$, a systematic study of axially symmetric
YM-Higgs configurations has revealed the existence of two possible types of asymptotics of the YM
fields, describing different ground states of the model.
These asymptotics are indexed by a number $m$, which is a positive integer.

\vspace{0.2cm}
{\bf Odd-$m$ conditions}

 For $m$ an odd number, $m=1,3,5,\dots$,
the YM fields possess a solution with
 \begin{eqnarray}
\nonumber
&&
H_1=0,~~H_2=1-m,
\\
\label{odd-k-flat}
&&
H_3=\frac{\cos \theta}{\sin \theta} \bigg(\cos ((m-1)\theta)-1\bigg) ,
\\
&&
\nonumber
H_4=-\frac{\cos \theta}{\sin \theta} \sin ((m-1)\theta),
\end{eqnarray}
which describe an infinite energy embedded Abelian configuration with a singular origin.
The configurations with these far field asymptotics carry a nonzero magnetic charge, $Q_M=n$
(with $Q_M$ computed
according to (\ref{Qm})).

However, in a flat spacetime background, a magnetically charged configuration
requires  Higgs fields to exist.
(or they are just embedded Abelian singular solutions).
Indeed, as discussed in \cite{Kleihaus:2003nj},
the Yang-Mills-Higgs (YMH) system possesses regular, finite energy solutions whose gauge potentials approach (\ref{odd-k-flat})
  in the far field.
  The better known case are the $m=1$, $n\geq 1$
  self-dual magnetic monopoles (which are in fact the only YMH closed form solutions).
  For $m>1$, they describe composite (non-self-dual)
 monopole-antimonopole configurations with a net magnetic charge.
A systematic study of these solutions can be found in \cite{Kleihaus:2003nj}.

 \vspace{0.2cm}
{\bf Even-$m$ conditions}


 A different picture is found for $m=2,4,\dots$.
The corresponding ground state YM solution reads
\begin{eqnarray}
\nonumber
&&
H_1=0,~~H_2=1-m,
\\
\label{even-k-flat}
&&
H_3=\frac{\cos ((m-1)\theta)-\cos \theta}{\sin \theta},
\\
\nonumber
&&
H_4=-\frac{\sin((m-1) \theta}{\sin \theta}.
\end{eqnarray}
Again, these asymptotics emerge from a systematic study of the axially symmetric YMH system \cite{Kleihaus:2003nj}.
The corresponding YMH solutions describe again (non-self dual) monopole-antimonopole chains.
However,  different from (\ref{odd-k-flat}), the total magnetic charge vanishes
in this case\footnote{Note that for any value of $m$,
the expression of the magnetic charge of the YMH solutions
found by employing the Abelian 't Hooft tensor
agrees with that from (\ref{Qm}).
}, $Q_M=0$.

Moreover,
as found in \cite{Ibadov:2004rt},
 in strong contrast to the odd$-m$ case,
these configurations survive in the limit of a vanishing Higgs field,
provided that the gravity effects are included.
For example, the well-known
Bartnik-McKinnon EYM particle-like solutions \cite{Bartnik:1988am}
are recovered for $m=2$, $n=1$.
The values  $m=2$, $n>1$ lead to their axially symmetric generalizations in \cite{Kleihaus:1997mn}.

One should also mention that,
as discussed in \cite{Kleihaus:2003nj},
the YM configuration (\ref{even-k-flat}) with $m=2k$
corresponds to a gauge transformed trivial solution,
\begin{eqnarray}
A_\mu =\frac{i}{e}(\partial_\mu U)U^\dagger.
\end{eqnarray}
However,
 the  $m=2k+1$ configurations (\ref{odd-k-flat}) describe a gauge transformed
 charge-$n$ Abelian multimonopole ($H_i\equiv 0$):
\begin{eqnarray}
A_\mu =\frac{i}{e}(\partial_\mu U)U^\dagger+U A_{\mu }^{(0)}U^\dagger,
\end{eqnarray}
where
\begin{eqnarray}
A_{\mu }^{(0)}=\frac{1}{2e}u_{\varphi}^{(n)}d\theta-n\sin \theta u_{\theta}^{(n)}d\varphi.
\end{eqnarray}
Also,
\begin{eqnarray}
 U={\rm exp}\{ -ik u_{\varphi}^{(n)} \},
 \end{eqnarray}
  for   both $m=2k+1$ and $m=2k$.

 \vspace{0.2cm}
{\bf The relation with the gauge field paramerization in \cite{Kleihaus:2003nj} and  \cite{Ibadov:2004rt}}

Although relying on the same ansatz proposed by Manton \cite{Manton:1977ht} and
Rebbi and Rossi \cite{Rebbi:1980yi},
the Ref. \cite{Kleihaus:2003nj}
expresses the YM potentials in a slightly different SU(2) basis,
\begin{eqnarray}
\label{v1}
 A_\mu dx^\mu=
\left( \frac{K_1}{r} dr + (1-K_2)d\theta\right)\frac{\tau_\vphi^{(n)}}{2e}
- n \sin\theta \left( K_3\frac{\tau_r^{(n,m)}}{2e}
                     + (1-K_4)\frac{\tau_\theta^{(n,m)}}{2e}\right) d\vphi~,
\end{eqnarray}
with the number $m$ entering also the SU(2) matrices:
\begin{eqnarray}
\nonumber
&&\tau_r^{(n,m)}=\sin m\theta (\cos n\varphi~\tau_x+\sin n\varphi~\tau_y)+\cos m\theta ~\tau_z,
\\
\nonumber
&&\tau_\theta^{(n,m)}=\cos m\theta (\cos n\varphi~\tau_x+\sin n\varphi~\tau_y)-\sin m\theta ~\tau_z,
\\
\nonumber
&&{~~~}\tau_\varphi^{(n)}=-\sin n\varphi~\tau_x+\cos n\varphi~\tau_y).
\end{eqnarray}
A direct comparison with (\ref{gauge-ansatz})
implies $H_1=K_1,~~H_2=K_2$, while
$
H_3=K_3\cos ((m-1)\theta)-(1-K_4)\sin((m-1)\theta),~~
1-H_4=K_3\sin ((m-1)\theta)+(1-K_4)\cos((m-1)\theta).
$

Note also that the Ref.  \cite{Ibadov:2004rt},
dealing with pure EYM solutions,
uses a version of (\ref{v1})
with $m\to k$ and  $1-K_4\to K_4$ (and a set of
boundary conditions at infinity resulting from (\ref{even-k-flat}), with $m=2k$).

We would like to emphasize that the description of a
given nA configuration in terms of
$(\tau_a^{(n,m)},K_i)$,
or  in terms of
$(u_a^{(n)},H_i)$
are equivalent.
The choice in this work for
$(u_a^{(n)},H_i)$
has the advantage to simplify somehow the emerging general picture,
providing a unified framework.
Also, it leads to slightly
 better numerical results for the gravitating solutions.

\subsubsection{YM far field asymptotics in AdS$_4$ spacetime}

As we know already from the
study in  \cite{Bjoraker:2000qd},  \cite{Winstanley:1998sn}
of the spherically symmetric case,
the $r\to \infty$ asymptotics of the YM fields
are less constrained for $\Lambda<0$.
In the generic axially symmetric case, an obvious condition results from the   requirement
that $T_t^t$, as given by (\ref{Ttt}), decays faster than $1/r^3$ as $r\to \infty$,
as imposed by the assumption that
the total mass $M$ is finite.
This implies
\begin{eqnarray}
F_{r\theta}^{(\varphi)} \to \frac{1}{r^{1+\epsilon}}, ~~~
F_{r\varphi}^{(r)} \to \frac{1}{r^{3+\epsilon}},~~~
F_{r\varphi}^{(\theta)} \to \frac{1}{r^{3+\epsilon}},~~~
F_{\theta\varphi}^{(r)} \to {\cal F}_{\theta\varphi}(\theta) ,~~
F_{\theta\varphi}^{(\theta)} \to {\cal G}_{\theta\varphi}(\theta) ,
\end{eqnarray}
with $\epsilon>0$ and
${\cal F}_{\theta\varphi}(\theta)$,
${\cal G}_{\theta\varphi}(\theta)$
two regular functions which vanish as $\theta\to 0,\pi$
as imposed by the regularity of the configurations.

A general expansion of the YM potentials $H_i$ in AdS spacetime reads
(with $i=1,\dots,4$):
\begin{eqnarray}
\label{gen-rel-AdS}
H_i= H_i^{(0)} +\frac{H_i^{(1)} }{r}+\frac{H_i^{(2)} }{r^2}+\dots,
\end{eqnarray}
where the functions $H_i^{(k)}$
depend on the angular coordinate $\theta$ only.
Once an expression is chosen for $H_i^{(0)}$, $H_i^{(1)}$
the functions $H_i^{(k)}$ (with $k\geq 2$)
are found by solving the YM equations
in the far field, as a power series in $1/r$.
Note, however that
the gauge condition (\ref{gauge})
implies
\begin{eqnarray}
 H_1^{(0)} =0,~~ H_2^{(0)} =const.,
\end{eqnarray}
while no obvious expressions are found for $H_3^{(0)}$, $H_4^{(0)}$
(note that the regularity of the solutions implies
$ H_2^{(0)}\big |_{\theta=0,\pi} =H_4^{(0)}\big |_{\theta=0,\pi}$).

In what follows, among all possible sets,
we shall restrict ourselves
to those YM asymptotics
which provide natural AdS generalizations of
the flat spacetime boundary conditions (\ref{odd-k-flat}), (\ref{even-k-flat}).


\vspace{0.2cm}
{\bf Odd-$m$ conditions}

One can prove that (\ref{odd-k-flat}) is still a solution of the
YM equations for $\Lambda<0$.
However, in strong contrast to the (asymptotically) flat case,
solutions with a non-zero magnetic charge
exist even in the absence of a Higgs field and/or
the gravity effects.
This can easily be seen
for the simplest case with $m=1,~n=1$,
where the following spherically symmetric exact solution is known:
\begin{eqnarray}
\label{ex-sol}
 H_1 =H_3=0,~~ H_2 =H_4=w(r)=\frac{1}{\sqrt{1+\frac{r^2}{\ell^2}}}.
\end{eqnarray}
This is the exact solution found in \cite{BoutalebJoutei:1979va}, which describes
a unit charge magnetic monopole with a finite mass $M ={3\pi^2}/{2e^2\ell}$.
However, the results in
\cite{Hosotani:2001iz}
show the existence of (numerical) generalizations of this solution
with the same behaviour at the origin, $w(0)=1$, and
  a relaxed set of boundary conditions at infinity,
 $ w(\infty)=w_0\neq 0$.
As discussed in \cite{Radu:2001ij},
these solutions admit axially symmetric generalizations with $(m=1,n>1)$,
possessing many similar properties.

A systematic study of possible boundary conditions compatible with regularity and finite
energy requirements
led us to the following AdS natural generalizations of
 (\ref{odd-k-flat}):
\begin{eqnarray}
&&
\nonumber
H_1=0,~~H_2=1-m+w_0,
\\
\label{odd-k-AdS}
&&
H_3=\frac{\cos \theta}{\sin \theta} \bigg(\cos ((m-1)\theta)-1\bigg)+w_0\sin((m-1) \theta),
\\
\nonumber
&&
H_4=-\frac{\cos \theta}{\sin \theta} \sin ((m-1)\theta) +w_0\cos((m-1) \theta),
\end{eqnarray}
with $w_0$ a constant which is not fixed apriori\footnote{Note that (\ref{odd-k-AdS})
is not a solution of the YM equations, unless $w_0=0$.
It describes instead the leading order
behaviour of the YM potentials in the AdS spacetime, as given by $H_i^{(0)}$.},
the flat space boundary conditions being recovered for $w_0=0$.
For example, one takes
\begin{eqnarray}
\label{odd-k-AdS-1}
H_1=0,~~H_2=w_0,~~H_3=0,~~H_4=w_0,~~{\rm for}~~m=1,
 \end{eqnarray}
and
\begin{eqnarray}
\label{odd-k-AdS-2}
H_1=0,~~H_2=w_0-2,~~H_3=(w_0-1)\sin 2\theta,~~H_4=-2 \cos^2\theta+w_0\cos 2\theta,~~{\rm for}~~m=3.
 \end{eqnarray}

 For the general asymptotics (\ref{odd-k-AdS}),
 one finds
 $
 F_{\theta\varphi}^{(r)}\to (1-w_0^2)n \sin\theta \cos((m-1)\theta),~~
  F_{\theta\varphi}^{(\theta)} \to (1-w_0^2)n \sin\theta \sin((m-1)\theta),
 $
 which implies
\begin{eqnarray}
\label{Q-AdS-odd}
Q_M=n|1-w_0^2|.
 \end{eqnarray}
Therefore, the magnetic charge of the  AdS configurations with the generic asynmptotics (\ref{odd-k-AdS})
is no longer an integer, a feature which
occurs already in the spherically symmetric case.
However, one can see that magnetically neutral solutions are found for $w_0=\pm 1$.

\vspace{0.2cm}
{\bf Even-$m$ conditions}

The situation is somehow different for even values of $m$.
It turns out that (\ref{even-k-flat})  provides the only possible set of  boundary
conditions compatible with the condition of a vanishing
magnetic charge\footnote{We did not find AdS generalizations of
 (\ref{even-k-flat}) with $Q_M=0$,
which would be
similar to
the odd-$m$  conditions (\ref{odd-k-AdS}).
In our approach, we consider a deformation of  (\ref{even-k-flat}) with:
   \begin{eqnarray}
H_1^{(0)}=0,~~H_2^{(0)}=1-m-w_0,~~
H_3^{(0)}=\frac{\cos ((m-1)\theta)-\cos \theta}{\sin \theta}+w_0S_3(\theta),~~
H_4^{(0)}=-\frac{\sin((m-1) \theta}{\sin \theta}+w_0S_4(\theta),
 \end{eqnarray}
 with $w_0$ a constant.
 The YM equations are supplemented with the zero-magnetic charge condition
$F_{\theta\varphi}^{(r)}\to 0$,
$F_{\theta\varphi}^{(\theta)}\to 0$,
 as $r\to \infty$.
This
 results in two first order differential equations for $S_3(\theta)$ and $S_4(\theta)$,
 which have closed form solutions. However,
 the functions $S_3(\theta)$ and $S_4(\theta)$ fail to satisfy the regularity conditions
 on the symmetry axis.
 Thus we conclude that (\ref{even-k-flat})
 are the only boundary conditions compatible with the assumption
 of a zero-magnetic flux at infinity.}.

Of course, this does not exclude the existence of AdS deformations of
(\ref{even-k-flat}). However, they would
generically possess a nonzero magnetic charge.
For example, we have studied
in a systematic way $m=2k$ solutions with
\begin{eqnarray}
&&
\nonumber
H_1=0,~~H_2=1-m w_0,
\\
\label{even-k-new}
&&
H_3=w_0\frac{\cos ((m-1)\theta)-\cos \theta}{\sin \theta},
\\
\nonumber
&&
H_4=1-w_0-w_0 \frac{\sin((m-1) \theta}{\sin \theta},
\end{eqnarray}
behaviour as $r\to \infty$,
such that
the $\Lambda=0$ conditions (\ref{even-k-flat}) are recovered for $w_0=1$.
For example, one has
\begin{eqnarray}
\label{even-k-AdS-1}
H_1=0,~~H_2=1-2w_0,~~H_3=0,~~H_4=1-2w_0,~~{\rm for}~~m=2,
 \end{eqnarray}
and
\begin{eqnarray}
\label{even-k-AdS-2}
H_1=0,~~H_2=1-4w_0,~~H_3=-2 w_0\sin 2\theta,~~H_4= 1-4 w_0\cos^2\theta,~~{\rm for}~~m=4.
 \end{eqnarray}
 This reveals an interesting feature:
the lowest  polar winding number solutions
 $m=1$
 and
 $m=2$,
 are in fact identical,
 since (\ref{odd-k-AdS-1}) results from  (\ref{even-k-AdS-1})
 via the identification $(1-2w_0)^{(m=2)}\to w_0^{(m=1)}$.
 This is related to the existence of a continuum of solutions
 (in terms of $w_0$),
 which is a pure AdS property.
 The solutions with higher $m$
 satisfy different
 boundary conditions and are inequivalent\footnote{This can be seen
 by comparing gauge invariant quantities, $e.g.$ the energy density. }.

The even-$m$ solutions with the asymptotics (\ref{even-k-new}) carry a magnetic charge
\begin{eqnarray}
\label{Q-AdS-even}
Q_M=\frac{mn}{2} |(1-w_0)w_0|.
 \end{eqnarray}
 Also, let us note that in contrast to the $\Lambda=0$ case,
 both  (\ref{odd-k-AdS}) and (\ref{even-k-new})
 correspond to intrinsic nA configurations (for $w_0\neq 0$ and $w_0\neq 1$, respectively).

\section{The solutions in the probe limit}

In this section we shall establish the existence of finite energy, regular
 YM configurations
 approaching at infinity the asymptotics (\ref{odd-k-AdS}) and (\ref{even-k-new}), respectively.
 Different from the case of a (asymptotically) Minkowski spacetime
 background, such configurations exist
 already in the probe limit ($i.e.$
 when neglecting
 the backreaction of the matter fields on the spacetime geometry).
 This approximation
 greatly simplifies the problem
but retains most of the interesting physics. For example, the probe limit was implemented recently
to analyse properties of axially symmetric solutions of the $SU(2)$ Yang-Mills-Higgs theory
\cite{Kichakova:2012pm}.

The problem has an intrinsic length scale $\ell$
(we recall $\Lambda=-3/\ell^2$).
Without any loss of generality,
we fix $\ell=1$,
the total mass of solutions, as given by (\ref{mass}),
being expressed in units of $4\pi/(e^2\ell)$.
The numerical approach employed in this case is similar to those described in Section 4
for gravitating configurations and we shall not discussed it here\footnote{Note
that the typical numerical accuracy is better for pure YM solutions,
the typical numerical error estimates being on the order of $10^{-5}$.}.
We mention only that the solutions are found by directly
 solving the YM equations with a given set of boundary conditions.

\subsection{The results}
Fixing $\Lambda=-3$,
 the only input parameters of the problem are the integers
 $(m,n)$
and the continuous constant $w_0$, which enters the boundary conditions at infinity  
(\ref{odd-k-AdS}), 
(\ref{even-k-new}).
We have studied a large number of configurations with $m= 1, 3,4,5,6,7$ and $n=1,\dots,10$.
Then we conjecture the existence of
 YM solutions in AdS$_4$ background for any values\footnote{This contrasts with the
 $\Lambda=0$ gravitating YM solutions in \cite{Ibadov:2004rt},
 which do not cover the whole $(m,n)$ space.} of $(m,n)$.

The YM solutions are found
subject to the following boundary conditions:
\begin{eqnarray}
\label{origin}
 H_{1}|_{r=0}=H_{3}|_{r=0}=0,~~
H_{2}|_{r=0}=H_{4}|_{r=0}= 1,
\end{eqnarray}
at the origin, and
(\ref{odd-k-AdS}),
(\ref{even-k-new}) at
infinity.

Also, for solutions with parity reflection symmetry (the only
type we consider in this paper),
the boundary conditions at $\theta=0,\pi/2$ are
\begin{eqnarray}
H_1|_{\theta=0,\pi/2}=H_3|_{\theta=0,\pi/2}=0,~~
\partial_\theta H_2|_{\theta=0,\pi/2}
=\partial_\theta H_4|_{\theta=0,\pi/2}
=0,
\end{eqnarray}
(therefore we need to consider the solutions only in the region $0\leq \theta \leq \pi/2$).

Regularity of the solutions on the symmetry axis imposes also
\begin{eqnarray}
\label{gauge-cond1}
 H_2\big |_{\theta=0,\pi} =H_4\big |_{\theta=0,\pi},
\end{eqnarray}
a condition which is verified from the numerical output.

For given $(m,n)$, the solutions are found by varying the parameter $w_0$
which enters the boundary conditions at infinity.
As expected, we have found a continuum of solutions
in terms of $w_0$.

 It is not easy to extract some
general characteristic properties of the solutions, valid for every choice of the parameters $(w_0, m, n)$.
However, we have found that the functions $H_i$ present always a considerable angle-dependence,
except for the $m=1$ case
(there one finds
usually a small angular dependence for the potentials $H_2, H_4$).
We have also noticed that the angular dependence generally increases with $Q_M$.
The profiles of typical $m=1,3,4,5$ solutions are given in Figures 1-4,
both as 3D-plots (with $\rho=r\sin\theta$, $z=r\cos\theta$), and as a function
of the radial coordinate for several different angles.
There we show the gauge potentials $H_i$ together with the mass-energy density
as given by $T_t^t$.

Also, we have found that for all sets  $(m,n)$,
in the absence of backreaction,
 the solutions exist for a single interval in $w_0$,
  the mass of the solutions strongly increasing for large values of $|w_0|$.
One can see this  in Figures 5, 6 for
the spectrum of the  $m=1,3,4,5,6$ solutions.
There we plot the mass of the solitons in terms of the parameter $w_0$
which enters the far field asymptotics  (\ref{odd-k-AdS}) and (\ref{even-k-new})
(we recall that $w_0$ fixes the magnetic charge of solutions via $Q_M=n|1-w_0^2|$
for odd $m$, and $Q_M=\frac{mn}{2} |(1-w_0)w_0|$ for even $m$).
It is interesting to notice that
the $M(w_0)$ picture
(as shown in Figures 5, 6), resembles the behaviour of
non-gravitating Q-balls, whose frequency-mass diagram exhibits a similar pattern, see $e.g.$ \cite{Kleihaus:2005me}.
However, different from
the frequency in that case, we could not derive analytical bounds for
the parameter $w_0$ which enters the boundary conditions for the YM potentials.

As seen in Figures 5, 6,
one important difference between odd-$m>1$
and even-$m$ solutions is that the former configurations
are disconnected from the vacuum\footnote{Since the $m=1$ and $m=2$
configurations coincide, this feature occurs also for the $m=1$ case.}.
That is, solutions with $M=0$ are found for $m=2k$ only.
This can be understood heuristically as follows.
For any azimuthal winding number $n\geq 1$,
the solutions with an even $m$ are found by  varying the parameter $w_0$ in
(\ref{even-k-new}), starting with $w_0=0$.
However, in that case $w_0=0$
results in a similar set of boundary conditions at $r=0$ and at infinity.
The only solution we could find in this case 
corresponds to the trivial one, $H_1=H_3=0,~H_2=H_4=1$,
with $M=0$.
Then $m=2k$ solutions with a small value of $|w_0|$
would be just deformations of this ground state and would possess a small mass.
However, this does not hold for $m=2k+1>1$,
since any value of $w_0$ is
leading to a set of boundary conditions at infinity different from the
 the trivial solution $H_1=H_3=0,~H_2=H_4=1$ 
 (as set by the boundary conditions at $r=0$).
 As a result, the mass of the odd-$m>1$ solutions possesses always a nonzero
 minimal value.

An unexpected feature of the odd-$m\geq 4$ solutions with large enough $n$
is their non-uniqueness.
That it, three different solutions are found for given $m,n$ and some range of the parameter $w_0$.
This property is shown in Figure 6 for $m=4,6$. To illustrate this
behaviour,  we also exhibit in Figure 7 the energy isosurfaces of the $m=6$ configurations at $w_0=0.7$ and
the fixed value of the energy density $\epsilon=2.1$.

One can understand this pattern as a manifestation of the composite structure of the $m\geq 4$
configurations.
Indeed, from Figure 7 we can clearly see that a typical $m=6$ solution consists of three constituents,
each of them representing a $m=2$ soliton.
Similarly, a $m=4$ configuration
consists of two $m=2$ components. Since each of the $m=2$ components of the composite configuration
possesses a magnetic dipole moment \cite{Ibadov:2004rt}, whose magnitude increases with $n$,
the energy of the dipole-dipole interaction between the components becomes a significant
part of the total energy.
However this interaction energy can be both repulsive and attractive,
depending on the orientation of the dipoles.
Thus the lowest branch corresponds to the aligned triplet of
dipoles, when the forces are most attractive (Figure 7, right plot), and two other branches correspond to the
higher mass solutions with two other possible orientations of the triplet of dipoles.

Also, one can observe that,
for $m=2k$ and $w_0=1$ in  (\ref{even-k-new}),
one finds  zero magnetic charge configurations\footnote{As noticed already, such solutions exist also
for $m=2k+1$ and $w_0^2=1$.},
which are the direct AdS counterparts of the solutions in \cite{Ibadov:2004rt}.

 \newpage
\setlength{\unitlength}{1cm}
\begin{picture}(15,20.85)
\put(0,0.4){\epsfig{file=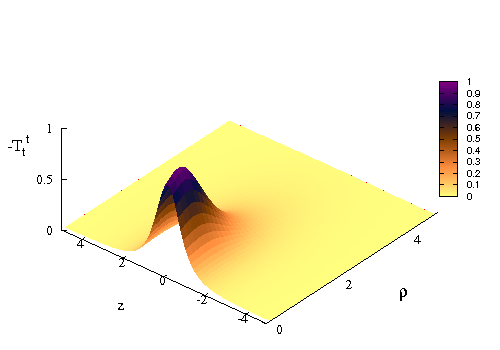,width=6.cm}}
\put(7,0.4){\epsfig{file=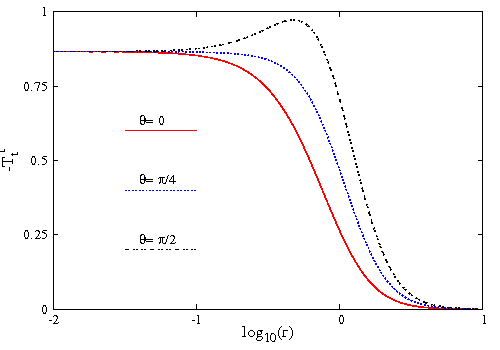,width=6.cm}}
\put(0,4.6){\epsfig{file=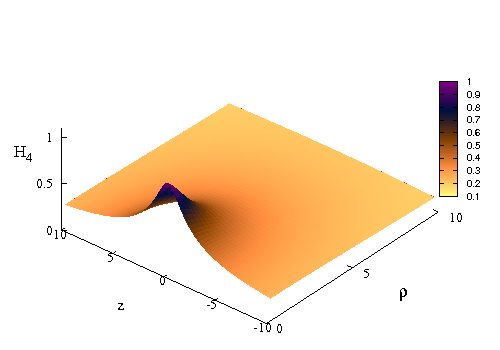,width=6.cm}}
\put(7,4.6){\epsfig{file=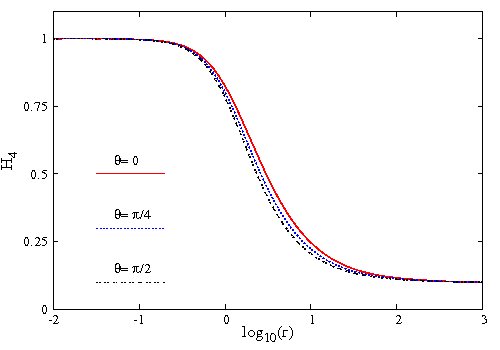,width=6.cm}}
\put(0,8.8){\epsfig{file=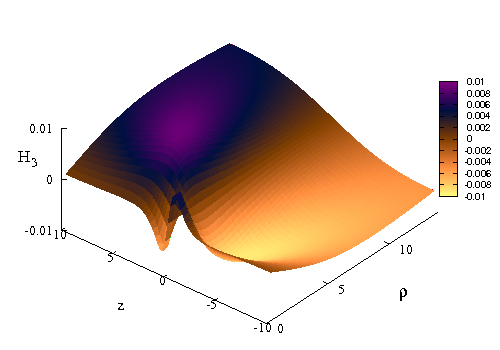,width=6.cm}}
\put(7,8.8){\epsfig{file=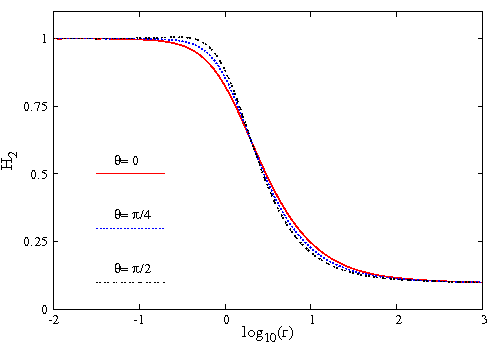,width=6.cm}}
\put(0,13.2){\epsfig{file=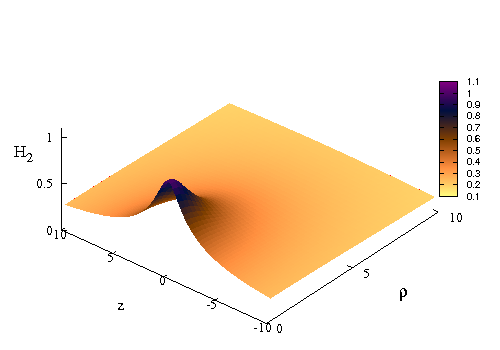,width=6.cm}}
\put(7,13.2){\epsfig{file=m1-H2-2d,width=6.cm}}
\put(0,17.6){\epsfig{file=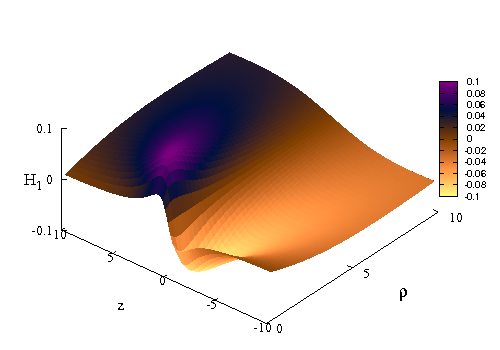,width=6.cm}}
\put(7,17.6){\epsfig{file=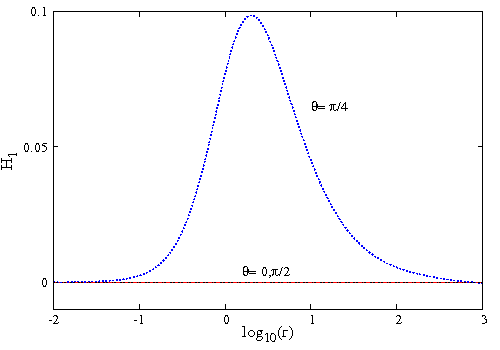,width=6.cm}}
\end{picture}
 \vspace*{ 0.cm}
\\
{\small {\bf Figure 1.} The profiles of a typical $m=1$ solution with $n=2$, $w_0=0.1$. }

 \newpage
\setlength{\unitlength}{1cm}
\begin{picture}(15,20.85)
\put(0,0.4){\epsfig{file=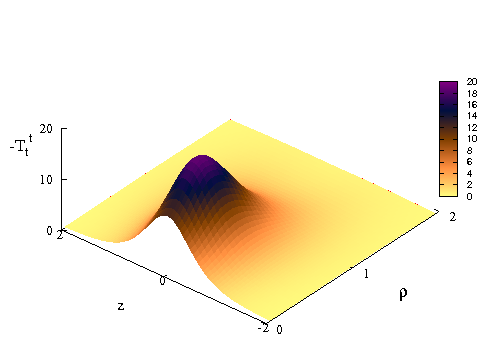,width=6.cm}}
\put(7,0.4){\epsfig{file=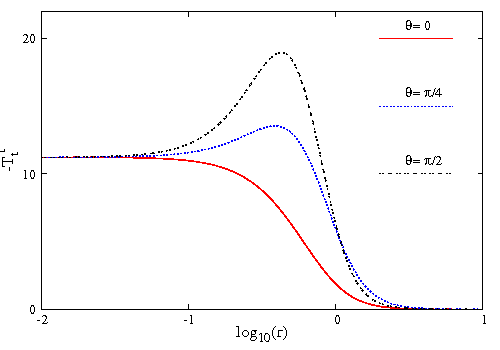,width=6.cm}}
\put(0,4.6){\epsfig{file=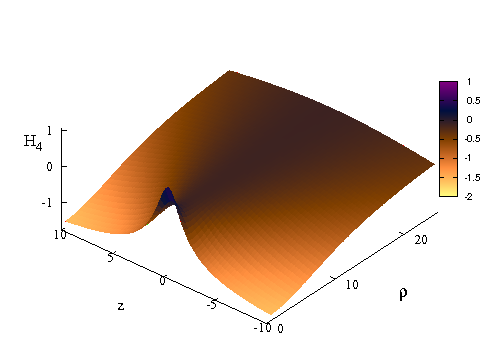,width=6.cm}}
\put(7,4.6){\epsfig{file=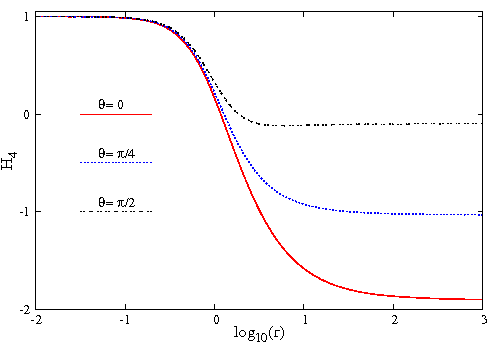,width=6.cm}}
\put(0,8.8){\epsfig{file=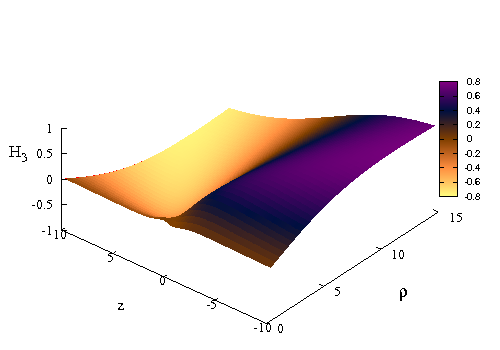,width=6.cm}}
\put(7,8.8){\epsfig{file=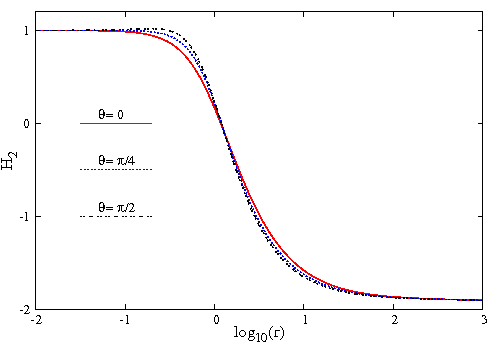,width=6.cm}}
\put(0,13.2){\epsfig{file=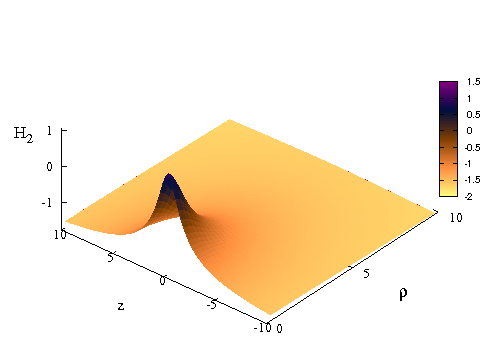,width=6.cm}}
\put(7,13.2){\epsfig{file=m3-H2-2d,width=6.cm}}
\put(0,17.6){\epsfig{file=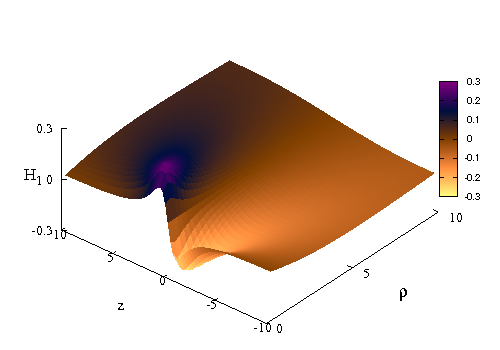,width=6.cm}}
\put(7,17.6){\epsfig{file=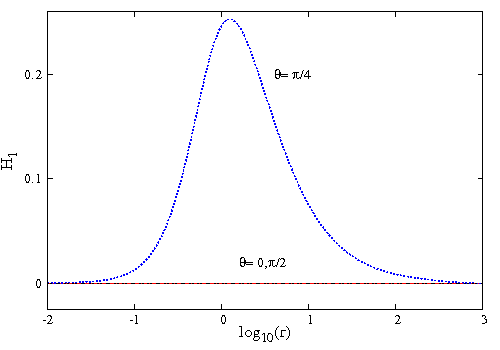,width=6.cm}}
\end{picture}
 \vspace*{ 0.cm}
\\
{\small {\bf Figure 2.} The profiles of a typical $m=3$ solution with $n=2$, $w_0=0.1$. }

 \newpage
\setlength{\unitlength}{1cm}
\begin{picture}(15,20.85)
\put(0,0.4){\epsfig{file=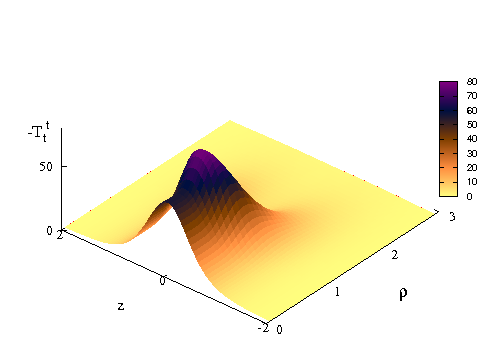,width=6.cm}}
\put(7,0.4){\epsfig{file=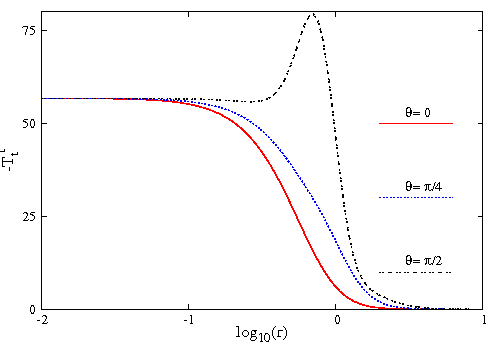,width=6.cm}}
\put(0,4.6){\epsfig{file=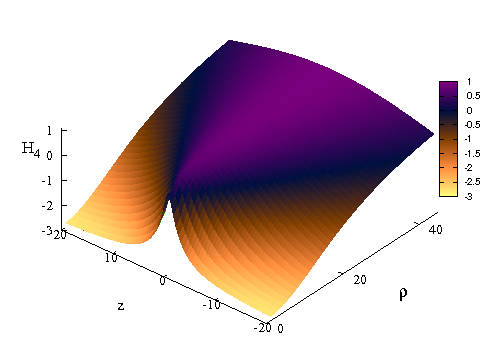,width=6.cm}}
\put(7,4.6){\epsfig{file=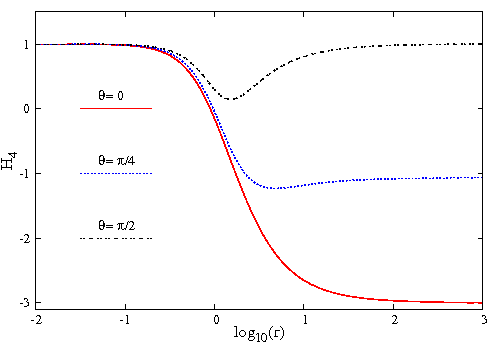,width=6.cm}}
\put(0,8.8){\epsfig{file=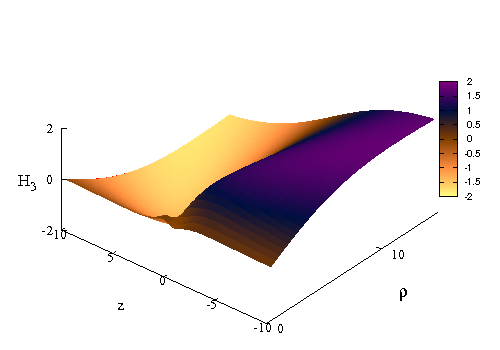,width=6.cm}}
\put(7,8.8){\epsfig{file=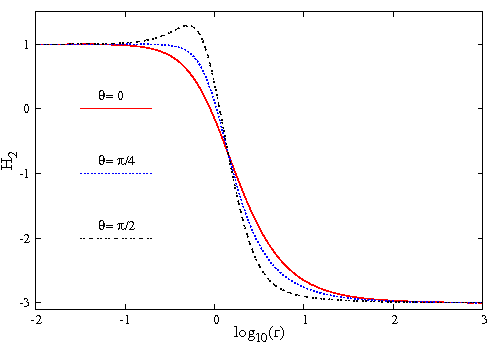,width=6.cm}}
\put(0,13.2){\epsfig{file=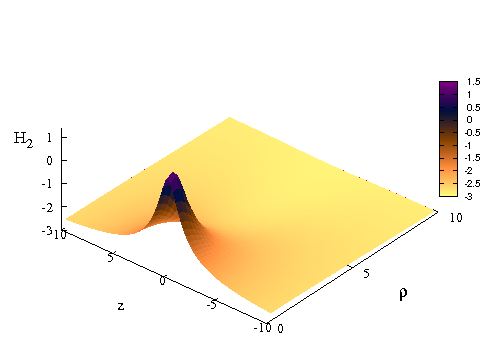,width=6.cm}}
\put(7,13.2){\epsfig{file=m4-H2-2d,width=6.cm}}
\put(0,17.6){\epsfig{file=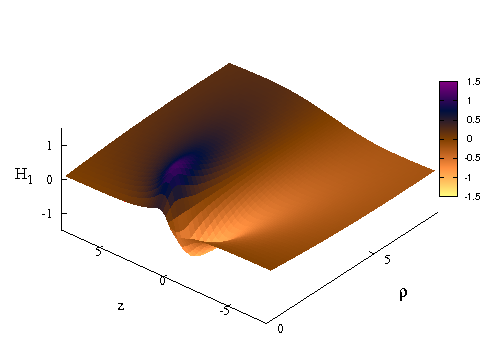,width=6.cm}}
\put(7,17.6){\epsfig{file=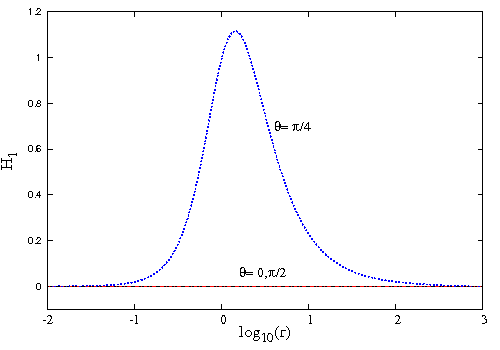,width=6.cm}}
\end{picture}
 \vspace*{ 0.cm}
\\
{\small {\bf Figure 3.} The profiles of a typical $m=4$ solution with $n=5$, $w_0=1$. }

 \newpage
\setlength{\unitlength}{1cm}
\begin{picture}(15,20.85)
\put(0,0.4){\epsfig{file=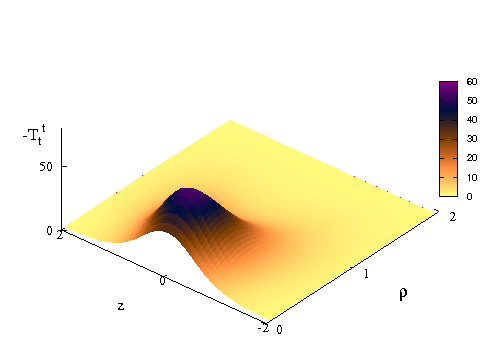,width=6.cm}}
\put(7,0.4){\epsfig{file=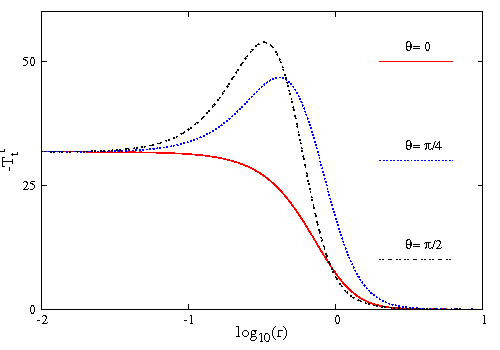,width=6.cm}}
\put(0,4.6){\epsfig{file=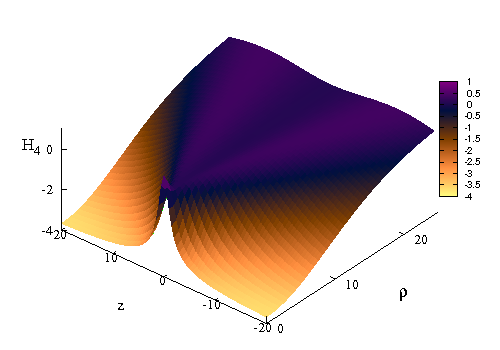,width=6.cm}}
\put(7,4.6){\epsfig{file=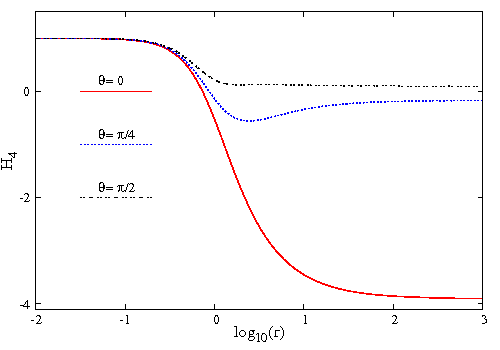,width=6.cm}}
\put(0,8.8){\epsfig{file=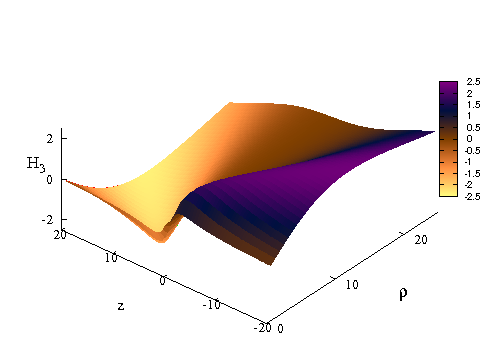,width=6.cm}}
\put(7,8.8){\epsfig{file=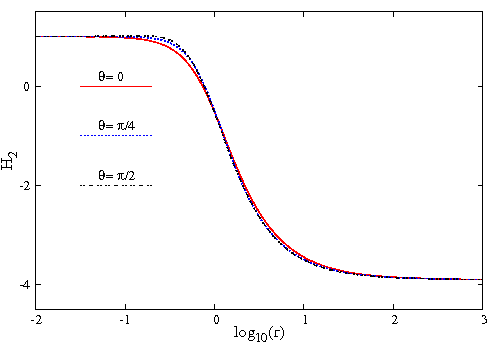,width=6.cm}}
\put(0,13.2){\epsfig{file=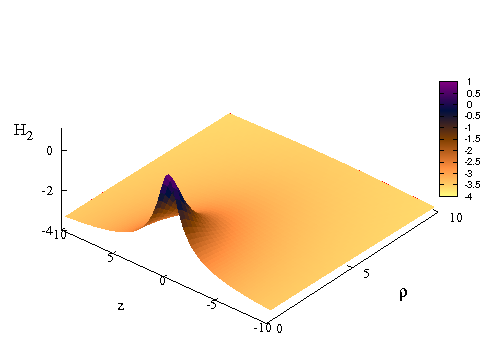,width=6.cm}}
\put(7,13.2){\epsfig{file=m5-H2-2d,width=6.cm}}
\put(0,17.6){\epsfig{file=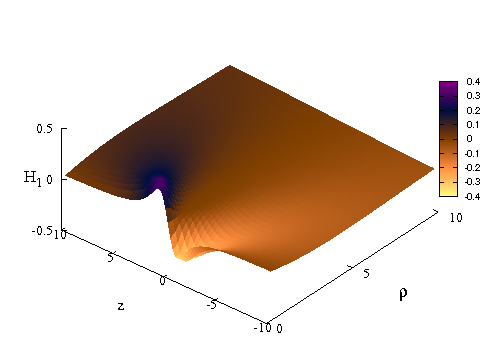,width=6.cm}}
\put(7,17.6){\epsfig{file=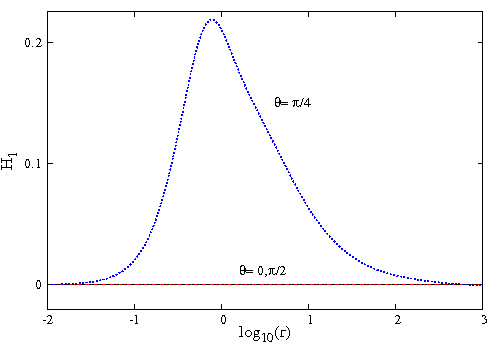,width=6.cm}}
\end{picture}
 \vspace*{ 0.cm}
\\
{\small {\bf Figure 4.} The profiles of a typical $m=5$ solution with $n=2$, $w_0=0.1$. }

\setlength{\unitlength}{1cm}
\begin{picture}(15,20.85)
\put(-0.5,10){\epsfig{file=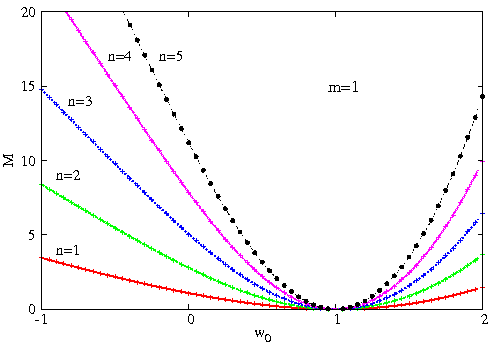,width=7.5cm}}
\put(8,10){\epsfig{file=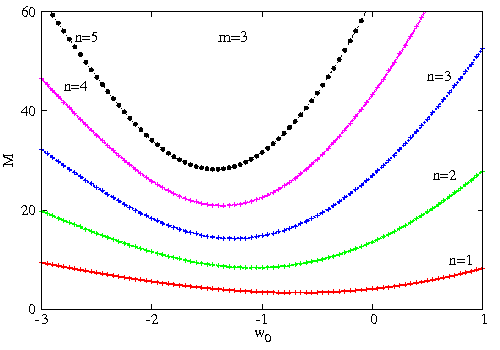,width=7.5cm}}
\put(-0.5,16){\epsfig{file=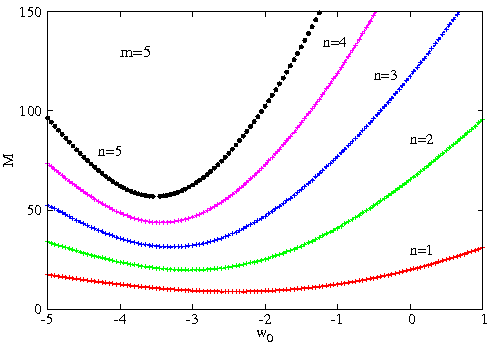,width=7.5cm}}
\put(8,16){\epsfig{file=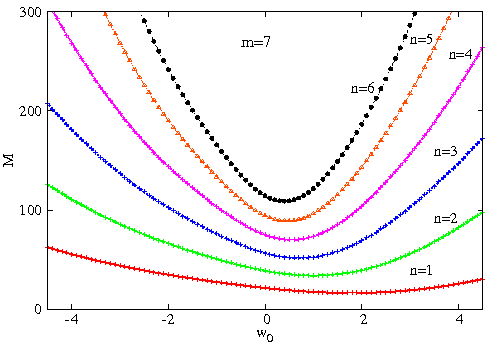,width=7.5cm}}
\end{picture}
\vspace*{-9.cm}
\\
{\small {\bf Figure 5.}
The spectrum of solutions is shown for  several odd-$m$ solutions.}

 \setlength{\unitlength}{1cm}
\begin{picture}(8,6.5)
\label{fig3}
\put(-0.5,0.0){\epsfig{file=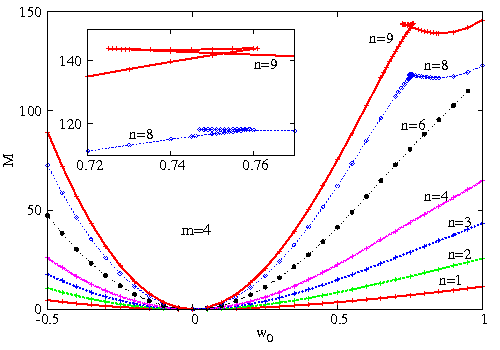,width=7.5cm}}
\put(   8,0.0){\epsfig{file=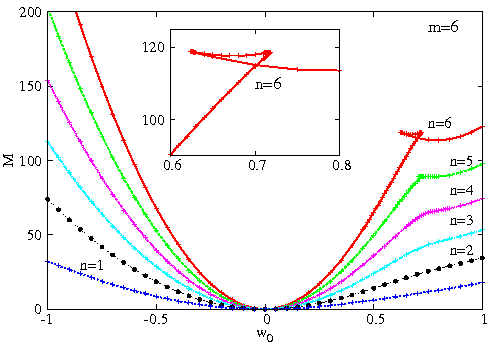,width=7.5cm}}\end{picture}
\\
\\
{\small {\bf Figure 6.} The same as Figure 5 for even-$m$ solutions.
One can notice the non-uniqueness of solutions for large enough values of
the winding number $n$.
}
\vspace{0.5cm}

 \setlength{\unitlength}{1cm}
\begin{picture}(8,6)
\label{fig7}
\put(-0.5,0.0){\epsfig{file=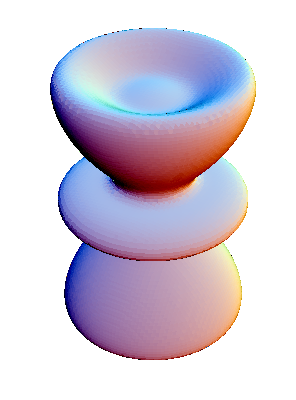,width=5cm}}
\put(   5,0.0){\epsfig{file=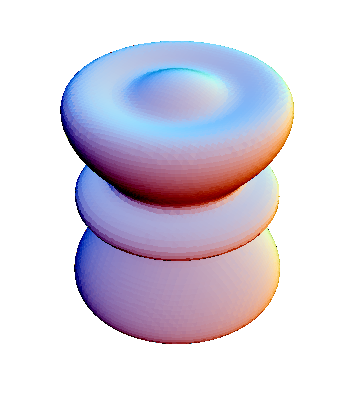,width=5cm}}
\put(   10,0.0){\epsfig{file=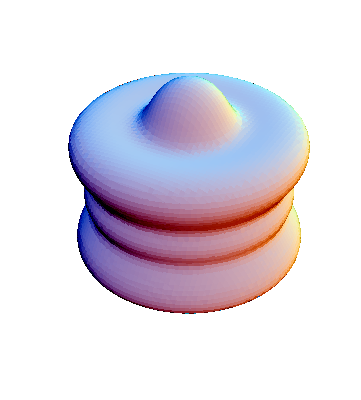,width=5cm}}
\end{picture}
\\
\\
{\small {\bf Figure 7.} The energy isosurfaces at the value of the energy density
$\epsilon=-T_t^t=2.1$ of the $m=6$, $n=6$ solutions are
shown for $w_0=0.7$.}
\vspace{0.5cm}
\\

\subsection{From AdS YM monopoles to flat space dynamical YM configurations}

We close this section by remarking that the  solutions discussed above
may be relevant in yet another direction.

It is well known that the
coordinate transformation (see, $e.g.$ \cite{Hawking})
\begin{eqnarray}
\label{mink1}
r=\frac{2R\ell^2}{\ell^2+T^2-R^2},~~
t=\ell
\left(
\arctan(\frac{R+T}{\ell})-\arctan(\frac{R-T}{\ell})
\right),
 \end{eqnarray}
puts the
 AdS metric (\ref{AdS})
in a conformally flat form
\begin{eqnarray}
\label{mink2}
ds^2=\frac{4\ell^2}{\ell^2+T^2-R^2)^2}
\big(dR^2+R^2(d\theta^2+\sin^2\theta d\varphi^2)-dT^2 \big).
 \end{eqnarray}
Then, due to the conformal invariance of the $d=4$ YM system,
any solution of the YM equations in an AdS background
results in a  solution in a Minkowski spacetime\footnote{
One can consider as well YM solutions in a fixed Einstein
universe background, see $e.g.$ \cite{Gibbons:1994du}.}.
For the case of the axially symmetric configurations discussed above,
these configurations are described by a YM ansatz
with a supplementary electric potential as compared to (\ref{gauge-ansatz}):
\begin{eqnarray}
\label{mink3}
 A_\mu dx^\mu
 &=&
\left( \frac{\bar H_1(R,\theta,T)}{R} dR + (1-H_2(R,\theta,T))d\theta\right)\frac{u_\vphi^{(n)}}{2e}
\\
\nonumber
&&
- n \sin\theta \left( H_3(R,\theta,T)\frac{u_r^{(n)}}{2e}
                     + (1-H_4(R,\theta,T))\frac{u_\theta^{(n)}}{2e}\right) d\vphi
                     +V(R,\theta,T) \frac{u_\vphi^{(n)}}{2e} dT,
 \end{eqnarray}
with the same functions $H_2,H_3,H_4$
and
\begin{eqnarray}
\label{mink4}
\bar H_1(R,\theta,T)=H_1\frac{\ell^2+T^2+R^2}{\ell^2+T^2-R^2},~~
V(R,\theta,T)=-\frac{2H_1}{\ell^2+T^2-R^2}.
 \end{eqnarray}
One can see the gauge potentials acquire a dependence
on the time coordinate $T$, via the transformation (\ref{mink1}).

Due to their numerical nature,
the study of the physical properties
of the corresponding Minkowski spacetime YM solutions is a difficult task,
which is beyond the purposes of this work.
Here we mention only that the AdS
exact solution (\ref{ex-sol})
corresponds to the flat spacetime meron configuration \cite{meron},
written in a special gauge
 \cite{BoutalebJoutei:1979va}.
 Thus we conjecture that at least the $m=1,n>1$
 AdS solutions
are likely to be relevant
to the subject of multi-merons.

\section{Including the backreaction: Einstein--Yang-Mills-$\Lambda$ solutions }

Now we shall address the question on how the YM solutions
discussed in the previous Section would deform the spacetime
geometry, by including the backreaction and solving the
full set of EYM equations.

 \subsection{The EYM action and field equations}
 We consider the Einstein-Yang-Mills action with a cosmological term:
\begin{eqnarray}
\label{model-EYM}
I_{bulk}= \int d^4 x\sqrt{-g}\left [
\frac{1}{16\pi G}\left(R-2\Lambda \right)
-{\rm Tr}
\big \{
\frac{1}{2}
F_{\mu\nu}F^{\mu\nu}
 \big \}
  \right].
\end{eqnarray}
 When including the backreaction, apart from the YM equations (\ref{feqA}),
  one solves also  the
Einstein equations,
\begin{eqnarray}
\label{Einstein-eqs}
E_{\mu\nu}=R_{\mu\nu}-\frac{1}{2}Rg_{\mu\nu}+\Lambda g_{\mu\nu}-8 \pi G T_{\mu \nu}=0~.
\end{eqnarray}

 This model has two length scales: the Planck one, $L_{Pl}=\sqrt{4\pi G}$,
and the cosmological one, $\ell$, which is fixed by the cosmological constant.
These two length scales define the dimensionless ratio
\begin{eqnarray}
\label{alpha}
\alpha=\frac{\sqrt{4\pi G}}{e \ell}.
\end{eqnarray}

 \subsection{EYM-$\Lambda$ system as a truncation of the ${\cal N}=4$ gauged supergravity}
For the generic EYM case,
the value of $\alpha$ is not fixed
but an input parameter of the theory.
However, among all values of the cosmological constant,
there is a case of special interest.
This corresponds to a consistent truncation of the
${\cal N}=4$ SO(4) gauged supergravity in $d=4$ \cite{Cvetic:1999au}.
Note that this supergravity model can be viewed
as a reduction of $d=11$ supergravity on $S^7$ \cite{Pope:1985bu}.
Thus these particular EYM-$\Lambda$
solutions have a higher-dimensional interpretation.

The bosonic sector of ${\cal N}=4$ SO(4) gauged supergravity
 contains two SU(2) fields $F_{\mu \nu}$ and ${\tilde F}_{\mu \nu}$,
a dilaton $\phi$ and an axion $\chi$.
In the conventions of \cite{Cvetic:1999au}, the Lagrangian density of the model reads
\begin{eqnarray}
\label{N=4}
{\cal L} &=& R  - \frac{1}{2} \partial_\mu\phi \partial^\mu \phi - \frac{1}{2}
e^{2\phi}\partial_\mu\chi \partial^\mu \chi - V(\phi,\chi)\,
- \frac{1}{2} e^{-\phi}\, {F_{ \mu \nu}^a}  F^{a \mu \nu} \nn\\
&& -\frac{1}{2} \fft{e^\phi}{1+\chi^2\, e^{2\phi}}\, {  \wtd
F^{a}_{\mu \nu}}  \wtd F^{a\mu \nu}\, -
\frac{1}{2\sqrt{-g}}\chi\,\epsilon_{\mu \nu \rho \sigma} F^{a \mu
\nu}  F ^{a\rho \sigma} +\frac{1}{2\sqrt{-g}} \frac{\chi\,
e^{2\phi}}{1+\chi^2\, e^{2\phi}}\, \epsilon_{\mu \nu \rho \sigma}
\wtd F^{a \mu \nu} \wtd F^{a\rho \sigma},
\end{eqnarray}
 where the
potential $V(\phi,\chi)$ is
\begin{eqnarray}
V(\phi,\chi) = -2e^2\, (4+ 2 \cosh\phi + \chi^2\, e^\phi).
\end{eqnarray}
It is easy to verify that
\begin{eqnarray}
\label{trunc}
\chi=\phi=0 ~~~{\rm and}~~~\tilde F_{\mu \nu}=F_{ \mu \nu},
\end{eqnarray}
is a consistent truncation of the general model (\ref{N=4}).
As a result, we end up with the EYM-$\Lambda$ Lagrangian
\begin{eqnarray}
\label{L1}
{\cal L} = R+12 e^2  - {F_{ \mu \nu}^a}  F^{a \mu \nu}
\end{eqnarray}
Working in conventions with  a length scale set by $\ell$ ($i.e.$, taking
in the numerics  $\Lambda=-3$),
it follows that the generic EYM-$\Lambda$
Lagrangian reduces to
 (\ref{L1}) for the
value of the coupling constant 
 \begin{eqnarray}
\alpha= \sqrt{2},
\end{eqnarray}
(note also that $e=1/\sqrt{2}$).
As a result, all EYM solutions with the above particular ratio between Planck and cosmological length scales
extremize also the action of the $d=11$
supergravity.
Then, by using the formulas in \cite{Cvetic:1999au}
with two equal gauge fields, $A^a = {\tilde A}^a$, one can uplift any such $d=4$ configuration
to eleven dimensions.
For example, the corresponding $d=11$ metric ansatz reads
\begin{eqnarray}
  ds^2_{11} = ds_4^2 + 4d\xi^2 +
\cos^2\xi \sum_a (\Theta^a-A^a_\mu dx^\mu)^2 + \sin^2\xi
\sum_a (\tilde{\Theta}^a-A^a_\mu dx^\mu)^2,
\end{eqnarray}
where $ds_4^2$ is the four dimensional line element,
and $(\Theta^a$, $\tilde{\Theta}^a)$ are SU(2) right
 invariant one forms on two 3-spheres $S^3 , \tilde{S}^3$.
 The corresponding expression of the $d=11$ matter fields can be found $e.g.$ in \cite{Pope:1985bu},
 and we shall not display them here.

 Also, note that, however, the solutions   discussed in this work  are generically not supersymmetric.
The interesting task of constructing  supersymmetric configurations would require a different approach\footnote{To our
knowledge, this task has not been considered yet in the literature.
The supersymmetric solutions in 
\cite{Chamseddine:2004xu}
have been found for a different truncation of the ${\cal N}=4$ SO(4) gauged supergravity model
than (\ref{trunc}).}.

\subsection{Solving the gravity field equations: the Einstein-De Turck approach}

\subsubsection{The metric ansatz}

The previous work  \cite{Radu:2001ij}
has solved the EYM field equations by using
 a standard approach as originally proposed in \cite{Kleihaus:1997mn} for asymptotically flat  solutions.
 There one employs a metric ansatz with three unknown functions, $f_i$ $(i=1,2,3)$  and
\begin{eqnarray}
\label{1}
ds^2=f_1(r,\theta)(\frac{dr^2}{N(r)}+r^2 d\theta^2)
+f_2(r,\theta)r^2\sin^2\theta d\varphi^2-f_0(r,\theta)N(r)dt^2,
\end{eqnarray}
(we recall $N(r)=1+\frac{r^2}{\ell^2}$).
The line-element (\ref{1}) is inspired by the one in \cite{Kleihaus:1997mn}
and  uses a  quasi-conformal choice of the gauge for the $(r,\theta)$-part
of the metric ($i.e.$ $g_{r\theta}=0$, $g_{\theta \theta}/g_{rr}=r^2N(r)$).
In this approach, one solves only a part of the full set of Einstein equations (\ref{Einstein-eqs}), namely
\begin{eqnarray}
E_r^r+E_\theta^\theta=0,~E_\varphi^\varphi=0,~E_t^t=0.
\end{eqnarray}
The remaining equations, $E_r^r-E_\theta^\theta$ and $E_r^\theta$
provide two constraints, which were used to test
 the numerical accuracy of the results.

Our choice in this work was
to construct the solutions by employing the Einstein-De Turck approach.
This approach has been proposed in  \cite{Headrick:2009pv}, \cite{Adam:2011dn}
 (see also \cite{Wiseman:2011by} for a review),
and been employed recently in the study of various asymptotically AdS configurations\footnote{Note, however,
that only configurations in a Poincar\'e patch of AdS  (and possibly with Abelian matter fields), have been considered so far
in the literature.}.
This scheme has the advantage of not fixing $apriori$ a metric gauge,
and leads to an overall better quality of the numerical results.

In this approach one solves the so called Einstein-DeTurck (EDT) equations
\begin{eqnarray}
\label{EDT}
R_{\mu\nu}-\nabla_{(\mu}\xi_{\nu)}=\Lambda g_{\mu\nu}+2 \alpha^2 T_{\mu\nu}~,
\end{eqnarray}
instead of  (\ref{Einstein-eqs}),
where $\xi^\mu$ is a vector defined as
\begin{eqnarray}
\label{csi}
\xi^\mu=g^{\nu\rho}(\Gamma_{\nu\rho}^\mu-\bar \Gamma_{\nu\rho}^\mu),
\end{eqnarray}
and $\Gamma_{\nu\rho}^\mu$ is the Levi-Civita connection associated to the
spacetime metric $g$ that one wants to determine.
Also, a  reference metric $\bar g$ is introduced, with
$\bar \Gamma_{\nu\rho}^\mu$ the corresponding Levi-Civita connection.
Solutions to (\ref{EDT}) solve the Einstein equations
iff $\xi^\mu \equiv 0$ everywhere on
${\cal M}$.
To achieve this,
we shall impose boundary conditions  which are compatible with
$\xi^\mu = 0$
on the boundary of the domain of integration.
Then, this should imply $\xi^\mu \equiv 0$ everywhere,
a condition which is verified from  the numerical output.

In this approach, we use  a metric ansatz
with two additional functions as compared to (\ref{1}),
\begin{eqnarray}
\nonumber
ds^2=f_1(r,\theta)\frac{dr^2}{N(r)}+S_1(r,\theta)(rd\theta+S_2(r,\theta)dr)^2
+f_2(r,\theta)r^2\sin^2\theta d\varphi^2-f_0(r,\theta)N(r)dt^2.
\end{eqnarray}
%
The obvious reference metric is AdS$_4$ spacetime with
\begin{eqnarray}
\label{backg}
S_1=f_1=f_2=f_0=1,~~S_2=0.
\end{eqnarray}

\subsubsection{The boundary conditions}
While the boundary conditions for the gauge potentials $H_i$ are similar to those used in
the probe limit (and we shall not give them again here),
the choice of the boundary conditions for the metric functions require special care.
These boundary conditions are found by constructing an approximate form
of the solutions on the boundary of the domain
of integration compatible with the requirement $\xi^\mu = 0$.
For example, as $r\to 0$, the solution reads
\begin{eqnarray}
\nonumber
&&f_1(r,\theta)=\frac{(c_1^2-1)c_2}{c_1+\cos2\theta}+f_{12}(\theta)r^2+\dots,~
f_2(r,\theta)=c_2(1+c_1)+f_{22}(\theta)r^2+\dots,~
f_0(r,\theta)=f_{00}+f_{02}(\theta)r^2+\dots,~
\\
\label{exp-r0}
&&
S_1(r,\theta)=c_2(1+c_1\cos 2\theta)+S_{12}(\theta)r^2+\dots,~~
S_2(r,\theta)=\frac{\sin 2\theta}{c_1+\cos 2\theta}+S_{22}(\theta)r^2+\dots,
\end{eqnarray}
where $c_1,c_2,f_{00}$ are parameters which result from the numerics.
This implies the following boundary conditions, which are
imposed in the numerics:
\begin{eqnarray}
\nonumber
\partial_r f_1\big|_{r=0}=\partial_r f_2\big|_{r=0}=\partial_r f_0\big|_{r=0}
=\partial_r S_1\big|_{r=0}=\partial_r S_2\big|_{r=0}=0.
\end{eqnarray}
%
The far field behaviour of the solutions is
\begin{eqnarray}
\label{far-field}
 &&
 f_0=1+\frac{f_{03}(\theta)}{r^3}+ O(1/r^4),~~f_1= O(1/r^4),~~
 f_2=1+\frac{f_{23}(\theta)}{r^3}+ O(1/r^4),~~
 \\
 &&
 \nonumber
 S_1=1+\frac{s_{13}(\theta)}{r^3}+ O(1/r^4),~~
 S_2= O(1/r^5),
\end{eqnarray}
with $f_{03}(\theta$, $f_{23}(\theta)$, $s_{13}(\theta)$
functions fixed by the numerics. These functions satisfy the relations
\begin{eqnarray}
\nonumber
f_{03}+f_{23}+s_{13}=0,~~\cos\theta (s_{13}-f_{23})=\sin\theta \frac{d}{d\theta}(f_{03}+f_{23}-s_{13}).
\end{eqnarray}
Thus we impose
\begin{eqnarray}
\nonumber
 f_1\big|_{r=\infty}=  f_2\big|_{{r=\infty}}=  f_0\big|_{r=\infty}=S_1\big|_{r=\infty}=1,~S_2\big|_{r=\infty}=0.
\end{eqnarray}
For $\theta\to 0$ one finds
\begin{eqnarray}
\nonumber
 &&f_i=f_i^0(r)+f_i^2(r)\theta^2+\dots,~~(i=1,2,3),~~
 \\
 &&S_1=S_1^0(r)+S_1^2(r)\theta^2+\dots,~~S_2=S_2^1(r)\theta+ \dots,
\end{eqnarray}
(a similar expression holds for $\theta\to\pi$).
 Then the
  boundary conditions at $\theta=0$ are
\begin{eqnarray}
\nonumber
\partial_\theta f_1\big|_{\theta=0}=\partial_\theta f_2\big|_{\theta=0}=
\partial_\theta f_0\big|_{\theta=0}=\partial_\theta S_1\big|_{\theta=0}=S_2\big|_{\theta=0}=0.
\end{eqnarray}
Moreover, we shall assume again that the solutions are
symmetric $w.r.t.$ a reflection in the
equatorial plane, which implies
\begin{eqnarray}
\nonumber
\partial_\theta f_1\big|_{\theta=\pi/2}=\partial_\theta f_2\big|_{\theta=\pi/2}=
\partial_\theta f_0\big|_{\theta=\pi/2}=\partial_\theta S_1\big|_{\theta=\pi/2}=S_2\big|_{\theta=\pi/2}=0.
\end{eqnarray}

\subsubsection{The holographic stress tensor and the mass of the solutions}

Our solutions should describe some  equilibrium states in the dual CFT, 
  defined in an Einstein universe in $d = 3$ dimensions, with a line element
  $ds^2=\ell^2(d\theta^2+\sin^2 d\theta^2)-dt^2$
(note that in the absence of a horizon, they do not
possess an intrinsic temperature $T$).
The SU(2) gauge symmetry of the bulk action corresponds to a global SU(2) symmetry
in the dual field theory.
Also, for the considered boundary conditions,
it is natural to work in a grand canonical ensemble,
the central quantity being the grand potential $W$.
In AdS/CFT, we identify $W$ with $T$
times the on-shell bulk action (in Euclidean signature).
 We thus analytically continue to Euclidean signature and compactify
the time direction with (an arbitrary) period $\beta=1/T$.
Note that since our solutions are static, the
analytical continuation $t\to i \tau$
 has no effects at the level of the equations of motion,
and the solutions discussed above extremize also the Euclidean action.

As usual, the $I_{bulk}$ action (\ref{model-EYM}) is supplemented with a Gibbons-Hawking boundary term $I_{GH}$ \cite{Gibbons:1976ue}, 
and
a boundary counterterm $I_{ct}$ \cite{Balasubramanian:1999re}.  Both $I_{bulk}$ and $I_{GH}$ exhibit divergences,
which are canceled by  $I_{ct}$.
Following the standard prescription, we  introduce a hypersurface $r = r_0$ with
some large but finite $ r_0$.
Then we ultimately remove the regulator by
taking $r_0\to \infty$.
The leading terms in the asymptotic form
of the metric functions have been already computed above, Rel. (\ref{far-field}).
Then, by
using the equations of motion
(supposing a fixed point with $\xi^\mu=0$),
and the Killing identity $\nabla^\mu \nabla_\nu K _\mu=R_{\mu \nu}K^{\nu}$,
for the Killing vector $K^\mu=\delta_t^\mu$, one arrives at the final expression for the grand potential $W$
\begin{eqnarray}
\label{W-fin}
W=\frac{3}{8 G\ell^2}
\int_0^\pi d\theta \left(f_{23}(\theta)+s_{13}(\theta) \right)\sin \theta~.
\end{eqnarray}
Within the same approach, one can
construct a divergence-free boundary stress tensor according to the prescription in
\cite{Balasubramanian:1999re},
 by defining
\begin{eqnarray}
\label{s1}
 T_{ab}&=& \frac{2}{\sqrt{-h}} \frac{\delta I}{ \delta h^{ab}}
=\frac{1}{8\pi G}(K_{ab}-Kh_{ab}-\frac{2}{\ell}h_{ab}+\ell E_{ab}),
\end{eqnarray}
where $E_{ab}$ is the Einstein tensor of the induced boundary metric $h_{ab}$,
and $K_{ab}$ is the extrinsic curvature.

One finds in this way the following large-$r$ expressions of
the nonvanishing components of the boundary stress tensor:
\begin{eqnarray}
\label{boundary-Tij}
T_{\theta}^\theta=-\frac{3}{16\pi G \ell}\frac{(f_{03}(\theta)+f_{23}(\theta))}{r^3}+\dots,
~~
T_{\varphi}^\varphi = \frac{3}{16\pi G \ell}\frac{ f_{23}(\theta)}{r^3}+\dots,
~~
T_{t}^t= \frac{3}{16\pi G \ell}\frac{ f_{03}(\theta)}{r^3}+\dots.
\end{eqnarray}
Note that this is a  traceless tensor, which is a result expected from the AdS/CFT
correspondence, since even-dimensional bulk theories
 are dual to odd-dimensional CFTs that
have a vanishing trace anomaly.

From $T_{ab}$, one can define a
 mass of the solutions as the conserved charge associated with time translation
 symmetry of the induced boundary metric $h$ \cite{Balasubramanian:1999re}.
 A straightforward computation leads to the following expression:
\begin{eqnarray}
\nonumber
M=\frac{3}{8 G\ell^2}
\int_0^\pi d\theta (f_{23}(\theta)+s_{13}(\theta))\sin \theta,
\end{eqnarray}
which, as expected  in the absence of a horizon,
equals the grand potential $W$.


\subsection{Remarks on the numerics}

In the numerical construction of the solutions in this work
we have used a slight generalization of the
techniques
employed in \cite{Kleihaus:1997mn}, \cite{Ibadov:2004rt}
for
asymptotically flat EYM regular configurations.

Our numerical scheme can be summarized as follows.
In a first step, one chooses a
suitable combination of the  Einstein-De Turck--YM equations,
such that the differential equations for the metric and gauge functions are diagonal
in the second derivatives\footnote{Note that the equations for the metric
functions contain also terms with mixed derivative $\partial_{r\theta}$.} with respect to $r$
(one should remark that
the  Einstein-De Turck system is truly formidable (with some equations containing up to 300 terms),
and thus we shall not present it here).
Then the radial coordinate $r$ is compactified according to
\begin{eqnarray}
r=\frac{x}{1-x},
\end{eqnarray}
with $0\leq x\leq 1$, such that we do not   introduce a cutoff value $r_{max}$.
For the derivatives, this leads to the following substitutions at the level of the equations
\begin{eqnarray}
r F_{,r}   \longrightarrow  x (1-x) F_{,x} \ \ ,
~~
r F_{,r\theta}   \longrightarrow  x (1-x) F_{,x \theta} \ \ ,
~~
r^2 F_{,r,r}   \longrightarrow
x^2 \left( (1-x)^2 F_{,x,x}
  - 2 (1-x) F_{,x} \right)
\end{eqnarray}
for any function $F=(H_i;f_0,f_1,f_2,S_1,S_2)$.
In the next step, the  equations are  discretized on a ($x,~\theta$) grid with $N_x\times N_{\theta}$ points.
Usually,
the grid spacing in the $x-$direction is non-uniform, while the values of the grid points
in the angular direction are given by $\theta_k=(k-1)\pi/(2(N_{\theta}-1))$.
Typical grids have sizes around $150 \times 30$ points, a limit imposed by the computational constraints.

The resulting system is solved iteratively until convergence is achieved.
All numerical calculations have been
performed by using the professional package CADSOL,
 which uses a  Newton-Raphson finite difference method with
 an arbitrary grid and arbitrary consistency order
(a detailed description of  this package is given in \cite{schoen}).
 This code solves a given system of nonlinear partial differential equations
 subject to a set of boundary conditions on a rectangular domain.
 Apart from some initial guess for the  solution,
   CADSOL requires also
 the Jacobian matrices for the equations $w.r.t.$ the unknown
 functions and their first and second derivatives, and 
the boundary conditions.
This software package provides also error estimates for each function,
which allows to judge the quality of the computed solution.
The typical  numerical error
for the solutions reported in this work is estimated to be of the order of $10^{-4}$
(also, the order of the difference formulae  was 6).

 Another independent test of the numerics is provided by the functions
 $(\xi^1,\xi^2)$
 as defined by (\ref{csi}), together with their norm.
 Also, to monitor
the numerical errors and test the convergence of our code, we have computed $R$, the Ricci scalar
 (we recall that we set $\ell=1$ which implies $R=-12$).
 The  results presented below have typically $|\xi|^2<10^{-8}$,
 while the maximal value for $|1+R/12|$ was around $10^{-4}$
 (note that the numerical errors are maximal close to $r=0$ and much smaller for large $r$).
 Also, as discussed below, the numerical accuracy deteriorates towards the limits of the $w_0$-interval,
 and the $m=3$ configurations have typically a lower accuracy.

\subsection{The solutions}

In the numerics, we have studied in a systematic way the solutions which are relevant for the case of the
${\cal N}=4$ SO(4) gauged supergravity model.
However, smaller values of $\alpha$
have  been considered as well (see the last part of this Section) to clarify the relation between the
YM
solutions in the fixed AdS background and the generalized Bartnik-McKinnon solutions in the asymptotically flat space
\cite{Ibadov:2004rt}.

For the case of main interest, $\alpha=\sqrt{2}$,
we have been able to establish the existence of gravitating generalizations
of some of the branches discussed in Section 3.
 There we have studied systematically solutions with $m=1,4$ and $n=1,2,3,4,5$
 as well as $m=3,~n=1$.
A number of configurations with $m=6$, $n=1$ have been also constructed.

 Since the profiles of the gauge potentials are rather similar to those
found in the probe limit, we shall not display them here.
The profiles of the metric functions of two typical $m=1,4$
solutions are shown in Figure 8.

For all considered cases,
the solutions always exist for a single $w_0$-interval only (see Figures 9, 10),
with $w_0^{min}<w_0<w_0^{max}$ (we recall that the parameter $w_0$ enters the boundary
conditions at infinity, fixing the magnetic charge).
However, when including
the backreaction, the allowed range of $w_0$
for which we could construct solutions
strongly decreases.

This can be understood in analogy with the simpler $m=1,~n=1$ spherically symmetric case
\cite{Bjoraker:2000qd},
\cite{Hosotani:2001iz}.
There, as noticed above, one single branch of non-gravitating solutions is found,
centered around $w_0=1$.
This branch survives when including the backreaction, in which case one
finds, however, a smaller size of the $w_0$-interval,
as compared to the probe limit\footnote{Note that
only nodeless solutions ($H_2=H_4=w(r)>0$)
are found for large enough $|\Lambda|$.}.
Moreover, the size of this interval decreases with $|\Lambda|$.

 \newpage
\setlength{\unitlength}{1cm}
\begin{picture}(15,20.85)
\put(0,0.4){\epsfig{file=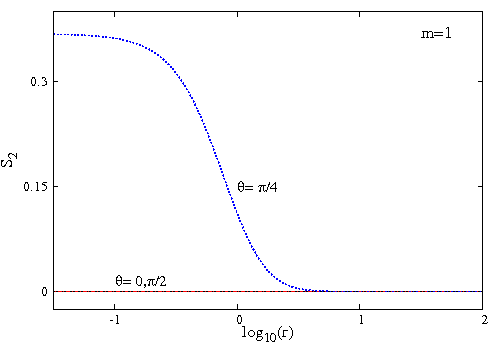,width=6.cm}}
\put(7,0.4){\epsfig{file=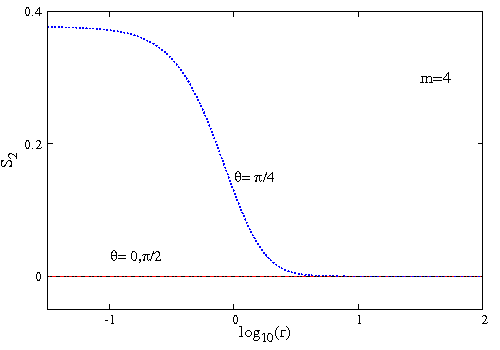,width=6.cm}}
\put(0,4.6){\epsfig{file=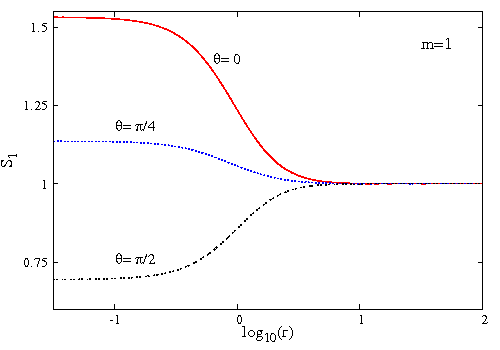,width=6.cm}}
\put(7,4.6){\epsfig{file=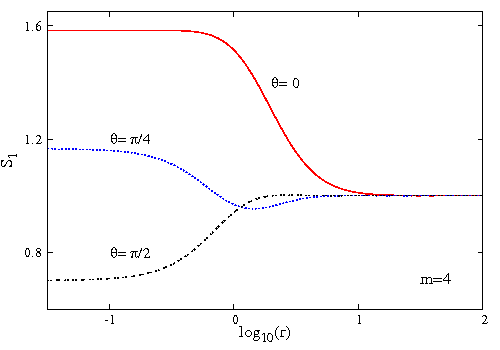,width=6.cm}}
\put(0,8.8){\epsfig{file=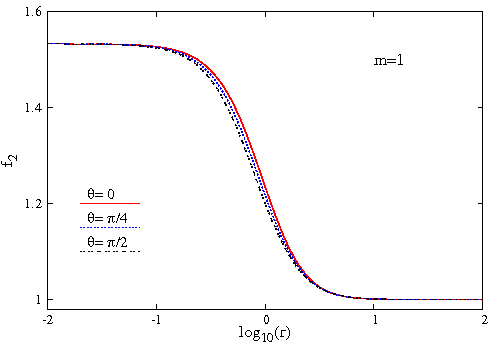,width=6.cm}}
\put(7,8.8){\epsfig{file=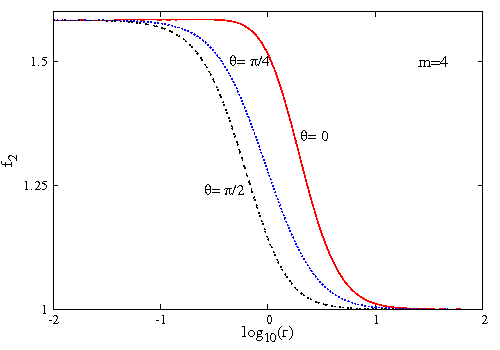,width=6.cm}}
\put(0,13.2){\epsfig{file=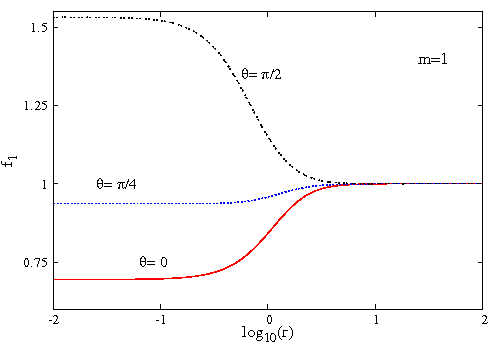,width=6.cm}}
\put(7,13.2){\epsfig{file=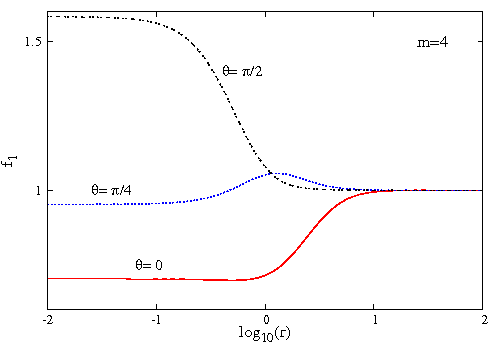,width=6.cm}}
\put(0,17.6){\epsfig{file=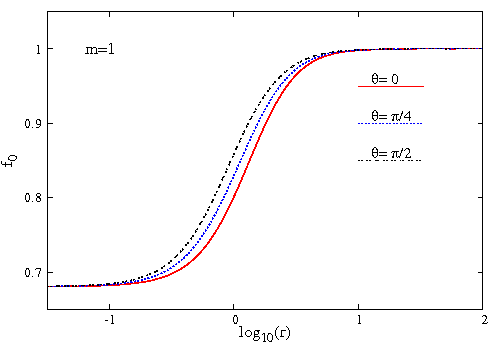,width=6.cm}}
\put(7,17.6){\epsfig{file=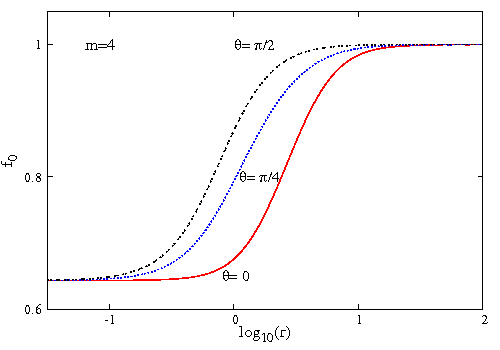,width=6.cm}}
\end{picture}
 \vspace*{ 0.cm}
\\
{\small {\bf Figure  8.}
The profiles of the metric functions for typical $\alpha=\sqrt{2}$,
  $m=1,4$ solutions
with  $n=4$, $w_0=0.75$ and  $n=3$, $w_0=-0.175$, respectively. }

\newpage
\setlength{\unitlength}{1cm}
\begin{picture}(8,6)
\put(-0.5,0.0){\epsfig{file=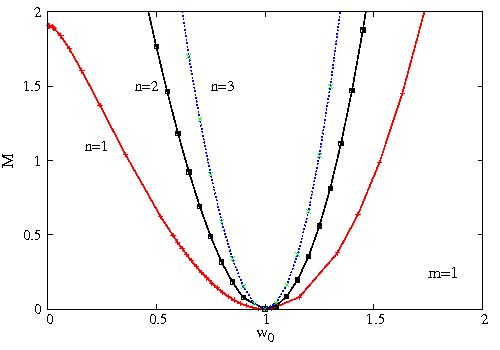,width=8cm}}
\put(8.2,0.0){\epsfig{file=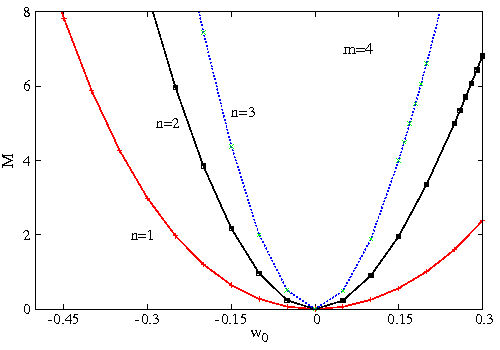,width=8cm}}
\end{picture}
\\
\\
{\small {\bf Figure 9.}
The spectrum of  $m=1,4$ magnetically charged  solutions of the
  ${\cal N}=4$ SO(4) gauged supergravity model.}
\\
\\
\setlength{\unitlength}{1cm}
\begin{picture}(8,6)
\put( 0.,0.0){\epsfig{file=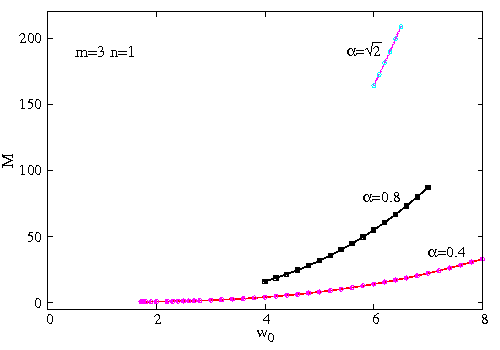,width=8cm}}
\put(8.7,0.0){\epsfig{file=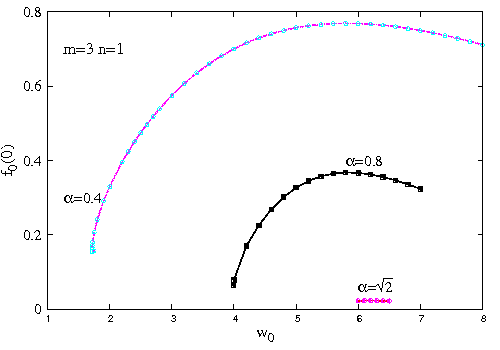,width=8cm}}
\end{picture}
\\
\\
{\small {\bf Figure 10.} {\it Left:} The spectrum of  $m=3,~n=1$ magnetically charged gravitating solutions
with several values of the coupling constant $\alpha$.
{\it Right:} The value of the metric function $f_0=-g_{tt}$ at $r=0$
is shown for the same configurations.}
 \vspace*{0.7cm}
\\
As $w_0\to w_0^{min}$, an extremal black hole
is approached, with a finite mass and horizon area, while
the solutions diverge as $w_0\to w_0^{max}$ \cite{Breitenlohner:2003qj}.

However, as found in  \cite{Hosotani:2001iz},
new $w_0$-intervals emerge as $|\Lambda|$ becomes smaller
and the shape of the branches has approximate self-similarity.
When the parameter $\Lambda$ approaches zero,
an already-existing $w_0$-interval of monopole  solutions
collapses to a single point in the moduli space, $w_0=\pm 1$
being the only allowed values\footnote{Viewed from this perspective,
the fundamental AdS branch of solution
reduces to the one-node Bartnik-McKinnon solution.}.
This implies the existence of  a fractal structure in
the moduli space of the solutions  \cite{Hosotani:2001iz}.

The same pattern is likely to exist for other values of $(m,n)$
(although the study of the  solutions is much more difficult in this case).
To clarify this behaviour in the odd-$m$ case ($m>1$), we plot in  Figure 9 (left)
the spectrum of solutions
for $(m=3,n=1)$ configurations with several values of $\alpha$.
One can see that the $w_0$-interval
decreases with increasing $\alpha$.
At the same time,
the numerical construction of the solutions
becomes more and more intricate.
This can be partially understood
from Figure 10 (right),
where we show the value of the metric function $g_{tt}$ at $r=0$
(which is an invariant quantity).

 \setlength{\unitlength}{1cm}
\begin{picture}(8,6)
\label{fig9}
\put( 1.5,0.0){\epsfig{file=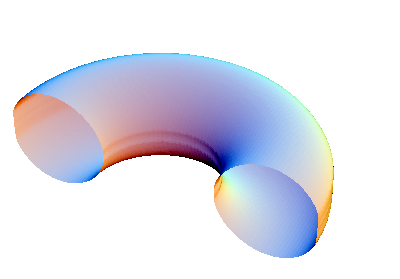,width=6cm}}
\put( 10,0.0){\epsfig{file=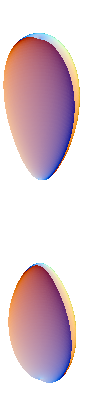,width=1.5cm}}
\end{picture}
\\
\\
{\small {\bf Figure 11.} Surfaces of constant energy density close to the maximum value of $-T_t^t$
are shown for typical $m=3$ (left) and $m=4$ (right) gravitating solutions.}
\vspace{0.5cm}
\\
One can see that, for any allowed $w_0$ and $\alpha=\sqrt{2}$,
the values of $f_0(0)$ are very small;
at the same time, the values at $r=0$
of the functions $f_1,~S_1$ (not shown there) are always very large.

This behaviour  makes the numerical construction of the $m=3$
solution within the ${\cal N}=4$ SO(4) gauged supergravity model
very difficult,
since the gravity effects are always very strong
in this case\footnote{Note that the numerical accuracy
of the reported $m=3$ gravitating solutions with $\alpha=\sqrt{2}$
is much lower as compared to all other solutions
in this work.
At the same time, the $m=3$ solutions with small enough $\alpha$
could be constructed with very good accuracy.}.
In particular, we could not clarify the issue of
limiting solutions at the ends of the $w_0$-interval.

However, the numerical construction of $m=4$ configurations
has proven to be much simpler.
Since the boundary value $w_0=0$
corresponds to the AdS background case,
these solutions possess a weak gravity regime (for small enough $|w_0|$).
However, the AdS background
becomes increasingly deformed as we increase  $|w_0|$,
with larger values of $M$.

As a new feature, we notice that for both $m=3$ and $m=4$,
the decrease of the size of the $w_0$-interval results in the
disappearance, for $\alpha=\sqrt{2}$,
of the solutions with a zero net magnetic charge, $Q_M=0$.
Therefore, no counterparts of the even-$m$ solutions in  \cite{Ibadov:2004rt}
exist for the case of the ${\cal N}=4$ SO(4) gauged supergravity model.
Moreover, the same applies also for $m=1$,
in which case, the branches of solitons cease to exist before approaching $w_0=-1$
(note that $w_0=1$ there corresponds to the vacuum case $M=0$).
We conclude that all solutions
of the ${\cal N}=4$ SO(4) gauged supergravity model
with excited nA fields,
possess a net magnetic charge.

However, this is a $\Lambda-$dependent feature, and the
 magnetically neutral solutions are recovered for large enough values of $\ell$
 (see the discussion below for the case $m=4$).

Also, we have found that for $m=1,3$
the energy of the solutions is located in a small region around the origin.
The energy density typically exhibits a
torus-like structure, with a maximum being located in the equatorial plane.
(This becomes apparent by considering
surfaces of constant energy density of, $e.g.$, half the respective
maximal value of the energy density.)
Moreover, a double-tori structure appears for large enough values of $n$.
However, this changes for higher values of $m$,
in which case the maximum of the energy density is typically located on the symmetry axis.
Thus these are composite configurations
with a dumbbell-like structure. 
Such configurations have two distinct components only, located symmetrically
with respect to the equatorial plane, 
and kept in balance due to the repulsive stresses of the YM fields.
These features are illustrated in Figure 11, where we plot surfaces
of constant
energy density
for $m=1,4$ gravitating solutions.

 \setlength{\unitlength}{1cm}
\begin{picture}(8,6)
\label{fig11}
\put(-0.5,0.0){\epsfig{file=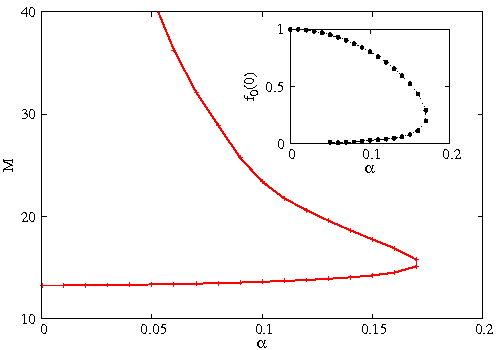,width=8cm}}
\put(   8,0.0){\epsfig{file=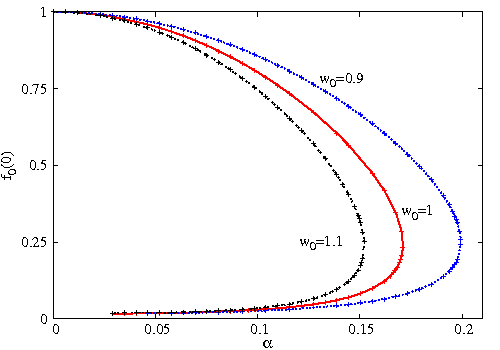,width=8.0cm, angle =-0}}\end{picture}
\\
\\
{\small {\bf Figure 12.}  {\it Left:} The values of the mass and the metric function $f_0(0)$  of the $(m=4,n=1)$
configurations
with zero magnetic charge $(w_0=1)$ are plotted as functions of   $\alpha$
  {\it Right:}
  The value  of the  metric function $f_0(0)$  of the $(m=4,n=1)$
configurations is plotted as a function of   $\alpha$
for several values of the parameter $w_0$.
  }
\vspace{0.5cm}

\subsection{EYM-$\Lambda$ configurations: from probe limit to
generalized Bartnik-McKinnon solutions}

An interesting pattern is observed,
once we abandon restrictions on the value of the parameters of the model
imposed by the reduction of the ${\cal N}=4$ SO(4) gauged supergravity.
In particular, we consider the dependence of
the solutions on the value of the coupling constant $\alpha$ (\ref{alpha}).
Furthermore, we may
restrict our consideration to the most interesting case $w_0 =0$ (for $m=2k+1$) and $w_0=1$ (for $m=2k)$,
in which case
 the boundary conditions are similar to those valid in the
asymptotically flat space.

Note that the dimensionless coupling constant $\alpha$ (\ref{alpha}) vanishes if
(i) the Newton constant $G \to 0$, or,
(ii) $\ell \to \infty$  (or, equivalently, $\Lambda \to 0$).
In the former
case one recovers
the  YM solutions in a fixed globally AdS background,
while the latter case
corresponds to the strongly coupled gravitating configurations in the asymptotically flat spacetime discussed in
\cite{Ibadov:2004rt}.

We have found numerical evidence that,
when gravity is coupled to the YM model in the globally AdS spacetime, a weakly gravitating solution emerges
from a particular  $(m,n)$ configuration which we discussed in the Section 3.
For these solitons the
values of the metric functions at small $\alpha$
are not very much different from those
corresponding to the background metric, see  (\ref{backg}).
However,
as the coupling constant $\alpha$ increases,
the background is more and more deformed.
At the same time, $\alpha$ cannot be arbitrarily large,
 the configurations approaching a critical
 solution  at some maximal value, $\alpha_{cr}$
 (which depends on $(m,n)$, see Figure 13).
There a backbending in $\alpha$ is observed,
with the occurrence of the second branch of solutions.
With decreasing $\alpha$, the solutions smoothly reach the corresponding (generalized) Bartnik-McKinnon  solutions in
\cite{Bartnik:1988am},
\cite{Kleihaus:1996vi},
\cite{Ibadov:2004rt},
 in the limit $\alpha \to 0$.
Along this branch, the mass of the $(m,n)$ configurations is higher
than the mass of the corresponding solution on the branch linked to the global AdS space.
Also, the
gravitational interaction remains strong and the metric function $f_0$ at the origin approaches
some very small but finite value, as seen in Figure 13.

We illustrate this pattern with a particular example of the $(m=4,~n=1)$ configurations presented in Figure 12 (left).
There we exhibit the evolution of the value of the mass and of the metric function $f(0)$ at the origin for the
$w_0=1$ solutions along both branches.
As seen in Figure 12 (right), a similar pattern is expected to hold for other values of $w_0$
(note also that the maximal (critical) value
of $\alpha$
decreases with $w_0$).

Now let us consider the matter and metric functions for small values of the rescaled coupling
constant $\alpha$. In Figure 13, top panel, we exhibit the gauge field
function $H_2$ and the metric function $f_0$ for $n=1$ configurations with $m=3,4,5,6$,
 on the upper branch
at $\alpha=0.05$. 

 \setlength{\unitlength}{1cm}
\begin{picture}(8,13)
\label{fig12}
\put(-1.0,11.8){\epsfig{file=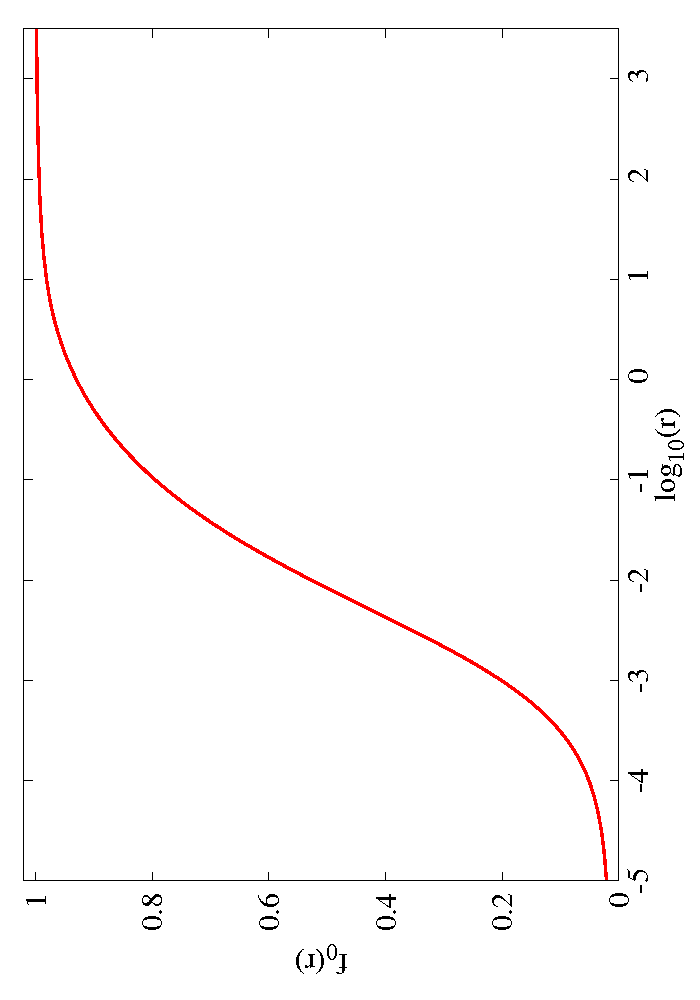,width=6cm, angle =-90}}
\put(   7.5,11.8){\epsfig{file=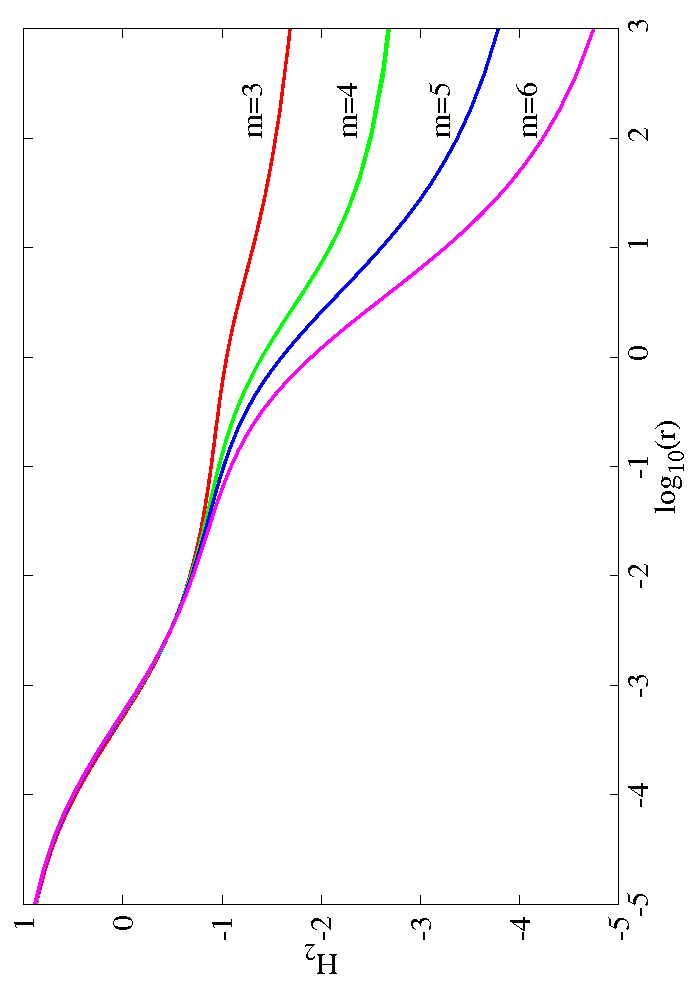,width=6.0cm, angle =-90}}
\put(3.5,-0.5){\epsfig{file=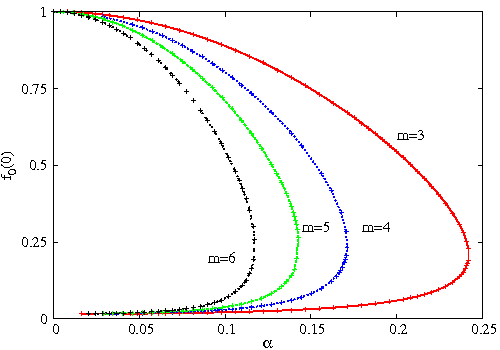,width=8.5cm, angle = 0}}
\end{picture}
\\
\\
\\
{\small {\bf Figure 13.}
  {\it Top plots:}
  The profiles of the metric function $f_0$ and the gauge function $H_2$ of the $n=1$, $m=3,4,5,6$,
configurations on the upper branches  are shown at $\alpha=0.05$.\\
{\it Bottom plot}: The values of the metric function $f_0(0)$  of the $n=1$
configurations with ($m=4,6$, $w_0=1$) and 
($m=3,5$, $w_0=0$) are plotted as functions of  $\alpha$.
  }
\vspace{0.5cm}

Clearly, there is an internal region where the gravitational interaction is strong and
the YM configuration agrees well with the first member of spherically symmetric Bartnik-McKinnon family of solutions
(in which case 
$H_1=H_3=0$,
$H_2=H_4=w(r)$,
with $w(r)$ interpolating monotonically between $1$ and $-1$). 
In the outer region
the metric functions rapidly approach their trivial values and the remaining $(m-2,1)$ configuration approaches to the
corresponding solution in the fixed AdS spacetime, which we considered above.

On the other hand, in the scaled coordinate  $r\to r/\alpha$, the inner region expands all the way up to infinity and contains the
generalized Bartnik-McKinnon solution with finite rescaled mass $\hat M = \alpha M$. Thus, the outer AdS configuration is
shifted up to infinity.

Note that this type of behavior is almost identical to the picture which holds for self-gravitating composite solitons
in the EYMH theory \cite{Kleihaus:2004fh,Kunz:2007jw}. In the latter case the gravitational coupling constant dependence
reveals a similar two-branch structure, however the branch lower in energy emerges from the corresponding composite YMH
solitons in the flat space \cite{Kleihaus:2003nj,Kleihaus:2004is}. At some maximal value of the coupling, this branch
bifurcates with the upper branch which, once again, extends all the
way back to the limit of vanishing coupling constant where it approaches the corresponding generalized Bartnik-McKinnon solution.
The difference to the EYM solitons in AdS spacetime is that in the EYMH model the evolution along the upper branch
 is related with the decrease of the expectation value of the Higgs field while in the former case it is related with the decrease
of  $|\Lambda|$.

\section{Further remarks. Conclusions}

The purpose of this work was to extend the spectrum of known
nA solutions in an AdS background by reporting
a whole new class of solitons indexed by two
integers $(m,n)$ and a continuous parameter $w_0$,
which fixes the magnetic charge of the solutions.
The main motivation for this task
comes from the observation that
the EYM action (\ref{model-EYM}) enters the $d = 4$ gauged supergravities as the
basic building block and one can expect the basic features of its solutions to be generic.

Our solutions are akin to the asymptotically flat
static, axially symmetric EYM configurations studied in \cite{Ibadov:2004rt}.
However, we have found that new features occur due to the different asymptotic
structure of the spacetime.
The most prominent one is perhaps the existence
in this case of a set of non-gravitating YM solutions
which,  for $\Lambda=0$,  are forbidden  in the absence of a Higgs field
(the odd-$m$ configurations, in the conventions of this work).
Also, the generic solutions possess a nonvanishing magnetic flux on the sphere at infinity,
being interpreted as non-topological monopoles.

New features occur also when including
 the backreaction of the YM fields on the spacetime geometry.
 A  rather unexpected result there is the
absence, for large enough values of $|\Lambda|$, of magnetically neutral configurations.
In particular, all solutions we have found for the case of main interest, corresponding to a consistent truncation
of  ${\cal N}=4$ SO(4) gauged supergravity, carry a non-zero magnetic charge.
 Another interesting result is the existence of balanced, regular composite configurations,
 with several distinct components.
 To our knowledge, this is the first example in the literature of AdS$_4$
 solutions with this property\footnote{It is interesting to note that
 the AdS$_4$ generalization of the Israel-Khan multi-black hole
 solution \cite{IsraelKhan}
 has not been found yet 
 (however, these configurations would possess conical singularities
for $\Lambda<0$  as well).
Moreover, the Majumdar-Papapetrou
\cite{Majumdar:1947eu},
\cite{Papaetrou:1947ib}
Einstein-Maxwell
extremal black holes are also not known
for a global AdS background (interestingly,
 such solutions exist for $\Lambda>0$ 
\cite{Kastor:1992nn},
\cite{Astefanesei:2003gw}).
}.

One question we did not address concerns the stability of the new solutions.
However,
it is known that some ($m= 1$, $n=1$) EYM solitons are  stable, and there is all reason to believe,
that at least some of the the new $(m, n)$ configurations are also stable.

There are various possible extensions of the solutions discussed in this work.
First, our preliminary results indicate the existence of
static axially symmetric
black hole solutions with nA hair.
These solutions are found within the same YM ansatz (\ref{gauge-ansatz}),
and the same boundary conditions at infinity,
by using again the EDT approach.
However, all configurations we have constructed so far
possess a single horizon of spherical topology.
Finding asymptotically AdS$_4$ multi-black hole solutions remains a
challenge\footnote{
In principle, one cannot exclude $apriori$ that
 $e.g.$ 
a composite EYM configuration with two solitons located symmetrically on the $z$-axis
would lead to a balanced dihole, when adding a small black hole at the center of each soliton.
}.

Another possible direction would be to  generalize these configurations
by including an electric potential for the YM connections.
These axially symmetric solitons would carry automatically
a nonzero electric charge  and a non-vanishing angular momentum density.

Finally, perhaps the most interesting task would be to generalize the solutions in this work
for a Poincar\'e patch of the  AdS spacetime.
In this case, all known solutions are just counterparts of the simplest
$(m=1,n=1)$
spherically symmetric
configurations in a globally AdS background.
We expect at least the $m=1$ configurations to
possess planar generalizations with an azimuthal
winding number $n>1$.

We hope to  report elsewhere on these problems.

 \vspace*{0.7cm}
\noindent{\textbf{~~~Acknowledgements.--~}}
We acknowledge  support by the DFG Research
Training Group 1620 �Models of Gravity�.
 E.R. gratefully acknowledges support from the FCT-IF programme and Ya.S.
support by the A.~von Humboldt Foundation in
the framework of the Institutes Linkage Programm.

\appendix

\setcounter{equation}{0}


\section{The spherically symmetric EYM-$\Lambda$ system: an exact solution}

The AdS generalizations of the $\Lambda=0$ Bartnik-McKinnon configurations
have been constructed  in \cite{Bjoraker:2000qd} by Bjoraker and Hosotani.
In such a study, it is convenient to employ Schwarzschild-like
coordinates, with a metric ansatz:
\begin{eqnarray}
\label{sph}
ds^2 = \frac{dr^2}{N(r)}
 + r^2 (d\theta^2 + \sin^2 \theta \, d\varphi^2)- N(r) \sigma^2(r) dt^2
\end{eqnarray}
where
\begin{eqnarray}
\nonumber
N(r) = 1 - \frac{2   m(r)}{ r}-\frac{\Lambda}{3} r^2 ,
\end{eqnarray}
$m(r)$ corresponding to the local mass-energy density.
The corresponding Yang-Mills ansatz is a truncation of
(\ref{gauge-ansatz}),
with $n=1$ and $H_1=H_3 =0$, $H_2=H_4=w(r)$.

Then the EYM equations take the simple form
\begin{eqnarray}
\nonumber
&&
m'=\alpha^2(\w'^2 N+\frac{(w^2-1)^2}{2 r^2}),
\\
&&
\label{eq-hoso}
w''+(\frac{N'}{N}+\frac{\sigma'}{\sigma})w'+\frac{w(1-w^2)}{r^2N}=0,
\\
\nonumber
&&
\sigma'=2\alpha^2\frac{\sigma w'^2}{r}.
\end{eqnarray}

The solutions with a regular origin
have the following behaviour at $r=0$
\begin{eqnarray}
\label{origin1}
w(r) = 1-br^2 +O(r^4),~~m(r)=2\alpha^2 b^2r^3+O(r^5),~~\sigma(r)=\sigma_0(1+4\alpha b^2r^2)+O(r^4),
\end{eqnarray}
where $b,~\sigma_0$ are real constants.
The corresponding expansion at large $r$ reads
\begin{eqnarray}
\label{asympt}
m(r)=M+ \frac{\alpha^2}{6r}(2\Lambda w_1^2-3(1-w_0^2)^2)+\dots,~
w(r) = w_0 +\frac{w_1}{r}+\frac{3w_0(1-w_0^2)}{2\Lambda r^2}+\dots,~
\sigma(r)=1-\frac{\alpha^2 w_1^2}{2r^4}+\dots,
\end{eqnarray}
where $w_0$, $M$ and $w_1$   are constants determined by
numerical calculations.
$M$ corresponds to the ADM mass of the
solutions, while
 $w_0$ determines the value of the magnetic charge, $Q_m= |1-w_0^2|$.

The numerical solutions are found by varying the  parameter $b$ which enters the expansion at the origin (\ref{origin1}).
Note that $b$ takes both positive and negative values,
 with $b=0$ corresponding to the ground (vacuum) state
\begin{eqnarray}
\label{ground}
w(r)=1,~m(r)=0,~\sigma(r)=1.
\end{eqnarray}

In this way
a continuum of monopole configurations is found,
which are regular in the entire space.
In particular, one finds solutions also for small ${b}$, which are arbitrarily close to the vacuum state
 ($i.e.$, $w(r)$ close to one everywhere).

This suggests to construct the configurations as  a power series expansion around the vacuum state (\ref{ground}),
%
 %
\begin{eqnarray}
\nonumber
w(r)=1+\sum_{k\geq 1}\epsilon^k w_k(r),~~m(r)=\sum_{k\geq 1}\epsilon^k m_k(r),
~~\sigma(r)=1+\sum_{k\geq 1}\epsilon^k \sigma_k(r),
\end{eqnarray}
with $\epsilon$ a small parameter.
To our best knowledge, although very simple, this approach did not appear in the literature.

To second order in $\epsilon$, the solution of the EYM equations reads
\begin{eqnarray}
\nonumber
&&
w_1(r)= 1-\frac{\ell}{r}\arctan\frac{r}{\ell} ,~~
w_2(r)=\frac{3 }{4}\frac{\ell}{r}\arctan\frac{r}{\ell}
\left((\frac{r}{\ell}+\frac{\ell}{r})\arctan\frac{r}{\ell}-1 \right),
\\
\label{ex-sol-n}
&&
m_1(r)=0,~~m_2(r)=\frac{\alpha^2 }{r}
\left(1-\frac{\ell}{r}\arctan\frac{r}{\ell}\right)
\left((\frac{r}{\ell}+\frac{\ell}{r})\arctan\frac{r}{\ell}-1 \right),
\\
\nonumber
&&
\sigma_1(r)=0,~~\sigma_2(r)=\frac{\alpha^2 }{2\ell^2}
\left (
\frac{3\pi^2}{4}-\frac{\ell^2}{r^2}(1+\frac{2}{1+\frac{\ell^2}{r^2}})
-\frac{\ell^4}{r^4}\arctan\frac{r}{\ell}
(
-\frac{2r}{\ell}+\frac{6r^3}{\ell^3}+(1+\frac{3r^4}{\ell^4})\arctan\frac{r}{\ell}
)
\right).
\end{eqnarray}

 \setlength{\unitlength}{1cm}
\begin{picture}(8,6)
\put(4,0.0){\epsfig{file=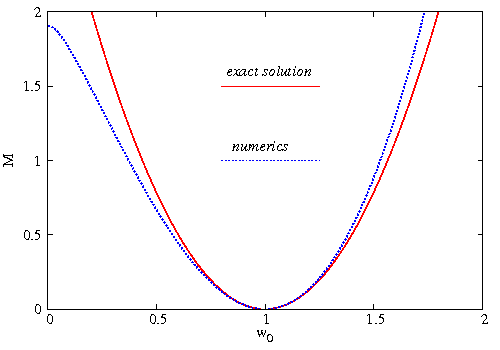,width=8cm}}
\end{picture}
\\
\\
{\small {\bf Figure 14.} The spectrum of spherically symmetric solutions
as given by the numerics and by the analytical estimates. }
\vspace{0.5cm}

Although one can extend the solution to higher orders,
it was not possible to identify a general
pattern\footnote{Note that the parameter $\epsilon$
 can be related to $b$ entering the small-$r$ expansion (\ref{origin1}),
 and also to $w_0$ which enters the
 asymptotics (\ref{asympt}).
 To second order, one finds $b=\frac{\epsilon(2+3 \epsilon)}{6\ell^2}$
 and $w_0=1+\epsilon  +\epsilon^2 \frac{3 \pi^2}{16}$, respectively.},
 the expressions for the functions 
 becoming increasingly complicated with increasing $k$.
 
 In Figure 14 we plot the spectrum of ${\cal N}=4$ SO(4)
 solutions as found by solving numerically the EYM equations
 together with the analytical estimates found from (\ref{ex-sol-n}).
 One can see that,  for
 small values of $ Q_M $, (\ref{ex-sol-n}) provides a reasonable approximation
 of the full numerical solutions.
One may expect similar closed form solutions to exist for other values of $(n,m)$ as well.

 \begin{small}
 
 \end{small}


\begin{thebibliography}{99}
\bibitem{BoutalebJoutei:1979va}
  H.~Boutaleb-Joutei, A.~Chakrabarti and A.~Comtet,
  Phys.\ Rev.\ D {\bf 20} (1979) 1884.
\bibitem{Volkov:1989fi}
  M.~S.~Volkov and D.~V.~Galtsov,
  JETP Lett.\  {\bf 50} (1989) 346
   [Pisma Zh.\ Eksp.\ Teor.\ Fiz.\  {\bf 50} (1989) 312];
\\
  H.~P.~Kuenzle and A.~K.~M.~Masood- ul- Alam,
  J.\ Math.\ Phys.\  {\bf 31} (1990) 928;
 \\
  P.~Bizon,
  Phys.\ Rev.\ Lett.\  {\bf 64} (1990) 2844.
\bibitem{Volkov:1994dq}
  M.~S.~Volkov and D.~V.~Galtsov,
  Phys.\ Lett.\ B {\bf 341} (1995) 279
  [hep-th/9409041].
\bibitem{Zhou:1991nu}
  Z.~h.~Zhou and N.~Straumann,
  Nucl.\ Phys.\ B {\bf 360} (1991) 180.

\bibitem{Wheeler}
R. Ruffini and J. A. Wheeler, Phys. Today 24 (1), 30 (1971).
\bibitem{Bekenstein:1996pn}
  J.~D.~Bekenstein,
  in ''{\it Moscow 1996, 2nd International A.D. Sakharov Conference on Physics}'', pp. 216-219
  [gr-qc/9605059].
\bibitem{Kleihaus:1997ic}
  B.~Kleihaus and J.~Kunz,
  Phys.\ Rev.\ Lett.\  {\bf 79} (1997) 1595
  [gr-qc/9704060].
\bibitem{Bartnik:1988am}
  R.~Bartnik and J.~Mckinnon,
  Phys.\ Rev.\ Lett.\  {\bf 61} (1988) 141.
\bibitem{Volkov:1998cc}
  M.~S.~Volkov and D.~V.~Gal'tsov,
  Phys.\ Rept.\  {\bf 319} (1999) 1
  [hep-th/9810070].
\bibitem{Winstanley:1998sn}
E.~Winstanley,
Class.\ Quant.\ Grav.\  {\bf 16} (1999) 1963.
\bibitem{Bjoraker:2000qd}
J.~Bjoraker and Y.~Hosotani,
Phys.\ Rev.\ D {\bf 62} (2000) 043513.


\bibitem{Sarbach:2001mc}
O.~Sarbach and E.~Winstanley,
Class.\ Quant.\ Grav.\  {\bf 18} (2001) 2125
[arXiv:gr-qc/0102033].
\bibitem{Breitenlohner:2003qj}
P.~Breitenlohner, D.~Maison and G.~Lavrelashvili,
Class.\ Quant.\ Grav.\  {\bf 21} (2004) 1667
[arXiv:gr-qc/0307029].

\bibitem{Baxter:2007at}
  J.~E.~Baxter, M.~Helbling and E.~Winstanley,
  Phys.\ Rev.\ Lett.\  {\bf 100} (2008) 011301
  [arXiv:0708.2356 [gr-qc]].
\bibitem{Baxter:2007au}
  J.~E.~Baxter, M.~Helbling and E.~Winstanley,
  Phys.\ Rev.\ D {\bf 76} (2007) 104017
  [arXiv:0708.2357 [gr-qc]].


\bibitem{Winstanley:2008ac}
  E.~Winstanley,
  Lect.\ Notes Phys.\  {\bf 769} (2009) 49
  [arXiv:0801.0527 [gr-qc]].
\bibitem{VanderBij:2001ia}
  J.~J.~Van der Bij and E.~Radu,
  Phys.\ Lett.\ B {\bf 536} (2002) 107
  [gr-qc/0107065].

\bibitem{Gubser:2008zu}
  S.~S.~Gubser,
  Phys.\ Rev.\ Lett.\  {\bf 101} (2008) 191601
  [arXiv:0803.3483 [hep-th]].
\bibitem{Gubser:2008wv}
  S.~S.~Gubser and S.~S.~Pufu,
  JHEP {\bf 0811} (2008) 033
  [arXiv:0805.2960 [hep-th]].
\bibitem{Manvelyan:2008sv}
  R.~Manvelyan, E.~Radu and D.~H.~Tchrakian,
  Phys.\ Lett.\ B {\bf 677} (2009) 79
  [arXiv:0812.3531 [hep-th]].
\bibitem{Ammon:2009xh}
  M.~Ammon, J.~Erdmenger, V.~Grass, P.~Kerner and A.~O'Bannon,
  Phys.\ Lett.\ B {\bf 686} (2010) 192
  [arXiv:0912.3515 [hep-th]].
\bibitem{Radu:2001ij}
  E.~Radu,
  Phys.\ Rev.\ D {\bf 65} (2002) 044005
  [gr-qc/0109015].
\bibitem{Radu:2004gu}
  E.~Radu and E.~Winstanley,
  Phys.\ Rev.\ D {\bf 70} (2004) 084023
  [hep-th/0407248].
\bibitem{Kleihaus:1997mn}
B.~Kleihaus and J.~Kunz,
Phys.\ Rev.\ D {\bf 57} (1998) 834.
\bibitem{Ibadov:2004rt}
  R.~Ibadov, B.~Kleihaus, J.~Kunz and Y.~Shnir,
  Phys.\ Lett.\ B {\bf 609} (2005) 150
  [gr-qc/0410091].
\bibitem{Majumdar:1947eu}
  S.~D.~Majumdar,
  Phys.\ Rev.\  {\bf 72} (1947) 390.
\bibitem{Papaetrou:1947ib}
  A.~Papaetrou,
  Proc.\ Roy.\ Irish Acad.\ (Sect.\ A)A {\bf 51} (1947) 191.
\bibitem{Cvetic:1999au}
  M.~Cvetic, H.~Lu and C.~N.~Pope,
  Nucl.\ Phys.\ B {\bf 574} (2000) 761
  [arXiv:hep-th/9910252].
\bibitem{Pope:1985bu}
C.~N.~Pope,
Class.\ Quant.\ Grav.\  {\bf 2} (1985) L77.

\bibitem{Manton:1977ht}
N.~S.~Manton,
Nucl.\ Phys.\ B {\bf 135} (1978) 319.
\bibitem{Rebbi:1980yi}
C.~Rebbi and P.~Rossi,
Phys.\ Rev.\ D {\bf 22} (1980) 2010.

\bibitem{Heusler:1996ft}
M.~Heusler,
Helv.\ Phys.\ Acta {\bf 69} (1996) 501.
\bibitem{Forgacs:1980zs}
P.~Forgacs and N.~S.~Manton,
Commun.\ Math.\ Phys.\ {\bf 72} (1980) 15,
\\
P.~G.~Bergmann and E.~J.~Flaherty,
J.\ Math.\ Phys.\ {\bf 19} (1978) 212.

\bibitem{Brihaye:1994ib}
  Y.~Brihaye and J.~Kunz,
  Phys.\ Rev.\ D {\bf 50} (1994) 4175
  [hep-ph/9403392].

\bibitem{Corichi:2000dm}
  A.~Corichi, U.~Nucamendi and D.~Sudarsky,
  Phys.\ Rev.\ D {\bf 62} (2000) 044046
  [gr-qc/0002078].

\bibitem{Kleihaus:2003nj}
  B.~Kleihaus, J.~Kunz and Y.~Shnir,
  Phys.\ Lett.\ B {\bf 570} (2003) 237
  [hep-th/0307110].



\bibitem{Kleihaus:2004is}
  B.~Kleihaus, J.~Kunz and Y.~Shnir,
  Phys.\ Rev.\ D {\bf 70} (2004) 065010
  [hep-th/0405169].

\bibitem{Chamseddine:2004xu}
  A.~H.~Chamseddine and M.~S.~Volkov,
  Phys.\ Rev.\ D {\bf 70} (2004) 086007
  [hep-th/0404171].

\bibitem{Hosotani:2001iz}
Y.~Hosotani,
J.\ Math.\ Phys.\  {\bf 43} (2002) 597.

\bibitem{Headrick:2009pv}
  M.~Headrick, S.~Kitchen and T.~Wiseman,
  Class.\ Quant.\ Grav.\  {\bf 27} (2010) 035002
  [arXiv:0905.1822 [gr-qc]].
\bibitem{Adam:2011dn}
  A.~Adam, S.~Kitchen and T.~Wiseman,
  Class.\ Quant.\ Grav.\  {\bf 29} (2012) 165002
  [arXiv:1105.6347 [gr-qc]].
\bibitem{Wiseman:2011by}
  T.~Wiseman,
  arXiv:1107.5513 [gr-qc].
\bibitem{Gibbons:1976ue}
  G.~W.~Gibbons and S.~W.~Hawking,
  Phys.\ Rev.\ D {\bf 15} (1977) 2752.
\bibitem{Balasubramanian:1999re}
V.~Balasubramanian and P.~Kraus,
Commun.\ Math.\ Phys.\  {\bf 208} (1999) 413
[arXiv:hep-th/9902121].

\bibitem{Kichakova:2012pm}
  O.~Kichakova, J.~Kunz, E.~Radu and Y.~Shnir,
  Phys.\ Rev.\ D {\bf 86} (2012) 104065
\bibitem{Hawking}
S.W. Hawking, G. F. R. Ellis,
\emph{The large structure of space-time},
Cambridge, Cambridge University Press, (1973).
\bibitem{Gibbons:1994du}
  G.~W.~Gibbons and A.~R.~Steif,
  Phys.\ Lett.\ B {\bf 346} (1995) 255
  [hep-ph/9412210].
\bibitem{meron}
V. de Alfaro, S. Fubini, G. Furlan, Phys.Lett. {\bf B65} 163 (1976).

\bibitem{Kleihaus:2005me}
  B.~Kleihaus, J.~Kunz and M.~List,
  Phys.\ Rev.\ D {\bf 72} (2005) 064002
  [gr-qc/0505143].
\bibitem{schoen}
 W. Sch\"onauer and R. Wei\ss ,
 J. Comput. Appl. Math. 27, 279 (1989) 279;
 \\
 M. Schauder, R. Wei\ss\ and W. Sch\"onauer,
 {\it The CADSOL Program Package},
 Universit\"at Karlsruhe, Interner Bericht Nr. 46/92 (1992).
\bibitem{IsraelKhan}
W. Israel and K. A. Khan, Nuovo Cimento {\bf 33} (1964) 331.
\bibitem{Kleihaus:1996vi}
  B.~Kleihaus and J.~Kunz,
  Phys.\ Rev.\ Lett.\  {\bf 78} (1997) 2527
\bibitem{Kleihaus:2004fh}
  B.~Kleihaus, J.~Kunz and Y.~Shnir,
  Phys.\ Rev.\ D {\bf 71} (2005) 024013
\bibitem{Kunz:2007jw}
  J.~Kunz, U.~Neemann and Y.~Shnir,
  Phys.\ Rev.\ D {\bf 75} (2007) 125008
\bibitem{Kastor:1992nn}
  D.~Kastor and J.~H.~Traschen,
  Phys.\ Rev.\ D {\bf 47} (1993) 5370
  [hep-th/9212035].
\bibitem{Astefanesei:2003gw}
  D.~Astefanesei, R.~B.~Mann and E.~Radu,
  JHEP {\bf 0401} (2004) 029
  [hep-th/0310273].
  
 \end{thebibliography}
\end{document}